\documentclass[twocolumn,aps,prl,floatfix,superscriptaddress]{revtex4-1}
\usepackage{amsmath,amssymb}
\usepackage[colorinlistoftodos, color=green!40, prependcaption]{todonotes}
\usepackage{graphicx,float}
\usepackage{times,txfonts}
\usepackage{textcomp}
\usepackage{color}
\usepackage{multirow}
\usepackage{color}
\usepackage{soul}
\usepackage{hyperref}
\usepackage{natbib}
\usepackage{nicefrac}
\usepackage{multirow}
\usepackage{blindtext}
\usepackage[export]{adjustbox}
\definecolor{hugoColor}{RGB}{59,134,255}

\begin{document}

\title{Tunable quantum emitters on large-scale foundry silicon photonics}

\author{Hugo~Larocque}
\affiliation{Research Laboratory of Electronics, Massachusetts Institute of Technology, Cambridge, Massachusetts 02139, USA}
\email{hlarocqu@mit.edu}

\author{Mustafa~Atabey~Buyukkaya}
\affiliation{Department of Electrical and Computer Engineering and Institute for Research in Electronics and Applied Physics, University of Maryland, College Park, Maryland 20742, USA}

\author{Carlos~Errando-Herranz}
\affiliation{Research Laboratory of Electronics, Massachusetts Institute of Technology, Cambridge, Massachusetts 02139, USA}
\affiliation{Institute of Physics, University of M\"unster, 48149, M\"unster, Germany}

\author{Samuel~Harper}
\affiliation{Department of Electrical and Computer Engineering and Institute for Research in Electronics and Applied Physics, University of Maryland, College Park, Maryland 20742, USA}

\author{Jacques~Carolan}
\affiliation{Research Laboratory of Electronics, Massachusetts Institute of Technology, Cambridge, Massachusetts 02139, USA}
\affiliation{Present Address: Wolfson Institute for Biomedical Research, University College London, London, UK}

\author{Chang-Min~Lee}
\affiliation{Department of Electrical and Computer Engineering and Institute for Research in Electronics and Applied Physics, University of Maryland, College Park, Maryland 20742, USA}

\author{Christopher~J.~K.~Richardson}
\affiliation{Laboratory for Physical Sciences, University of Maryland, College Park, Maryland 20740, USA}

\author{Gerald~L.~Leake}
\affiliation{State University of New York Polytechnic Institute, Albany, New York 12203, USA}

\author{Daniel~J.~Coleman}
\affiliation{State University of New York Polytechnic Institute, Albany, New York 12203, USA}

\author{Michael~L.~Fanto}
\affiliation{Air Force Research Laboratory, Information Directorate, Rome, New York, 13441, USA}

\author{Edo~Waks}
\affiliation{Department of Electrical and Computer Engineering and Institute for Research in Electronics and Applied Physics, University of Maryland, College Park, Maryland 20742, USA}

\author{Dirk~Englund}
\affiliation{Research Laboratory of Electronics, Massachusetts Institute of Technology, Cambridge, Massachusetts 02139, USA}

\begin{abstract}
Controlling large-scale many-body quantum systems at the level of single photons and single atomic systems is a central goal in quantum information science and technology. Intensive research and development has propelled foundry-based silicon-on-insulator photonic integrated circuits to a leading platform for large-scale optical control with individual mode programmability. However, integrating atomic quantum systems with single-emitter tunability remains an open challenge. Here, we overcome this barrier through the hybrid integration of multiple InAs/InP microchiplets containing high-brightness infrared semiconductor quantum dot single photon emitters into advanced silicon-on-insulator photonic integrated circuits fabricated in a 300~mm foundry process. With this platform, we achieve single photon emission via resonance fluorescence and scalable emission wavelength tunability through an electrically controlled non-volatile memory. The combined control of photonic and quantum systems opens the door to programmable quantum information processors manufactured in leading semiconductor foundries.
\end{abstract}

\maketitle

\section{Introduction}

Among single photon emitters (SPEs), III-V semiconductor quantum dots (QDs) stand out for near-unity internal quantum efficiency, purity, and indistinguishability~\cite{He:13, Somaschi:16, Ding:16, Wang:19, Senellart:17, Tomm:21, Thomas:21}, making them key building blocks in technologies requiring on-demand entangled photon pair emission~\cite{Dousse:10, Muller:14, Liu:19}, photon-photon interactions~\cite{Javadi:15, LeJeannic:21, LeJeannic:22}, or photonic cluster state generation~\cite{Lindner:09, Schwartz:16,Istrati:20,Cogan:23}. Recent advances in materials science and electron-nuclear spin control have also renewed interest in storing quantum information within these structures. Specifically, methods for reducing coupling between the QD electron spin with the nuclear spin bath of the embedding III-V material 
can push spin coherence times from tens of nanoseconds~\cite{Stockill:16} to beyond 0.1 ms~\cite{Gangloff:19, Gangloff:21,Zaporski:23}.


The central challenge now lies in developing systems for controlling many-QD quantum systems. This requires (i) efficiently mediated optical interactions enabled by low propagation losses and (ii) scalable control of individual QDs. Addressing (i) and (ii) simultaneously has motivated the development of photonic integrated circuit (PIC) platforms as on-chip solutions for this quantum information processing task. Successfully deploying such integrated quantum technologies to their full potential critically hinges on the compatibility of QD structures with a given PIC platform. Leading approaches include monolithic III-V PICs~\cite{Dietrich:16,Lodahl:18}, which have enabled interactions between multiple emitters~\cite{Grim:19, Papon:22, Tiranov:23}. However, optical attenuation in III-V waveguides exceeding 15~dB/cm~\cite{Jons:15} and the need for specialized manufacturing present obstacles on (i) and (ii), respectively. To address this constraint, recent work has pursued hybrid integration of quantum emitter materials with PIC platforms~\cite{Kim:20, Elshaari:20} including silicon~\cite{Kim:17, Katsumi:19, Katsumi:20} or silicon nitride~\cite{Mouradian:15,Zadeh:16, Elshaari:17, Davanco:17, Elshaari:18, Chanana:22}. However, none of these approaches combine (i) and (ii) with the scaling advantages of advanced foundry-based silicon-on-insulator (SOI) PIC platforms, which leverage general-purpose programmability, low power optical modulation, and the large-scale integration of thousands of individually controlled optical components.~\cite{Sun:13, Bogaerts:20, Bogaerts:20b, Edinger:21}. 

Here, we address this challenge by simultaneously realizing four key advances: (i) iterative design, manufacture, and post-processing methods of advanced SOI PICs with sub-3dB fiber coupling efficiency fabricated in a leading 300 mm foundry; (ii) the hybrid integration of multiple telecom quantum emitters via scalable transfer printing methods; (iii) resonance fluorescence from individually addressable SPEs; and (iv) the introduction of an electrically controlled non-volatile memory for individual SPE Stark shifting.

\section{Results}

\subsection{Architecture}

\begin{figure*}[t]
	\centering
	\includegraphics[width=\linewidth]{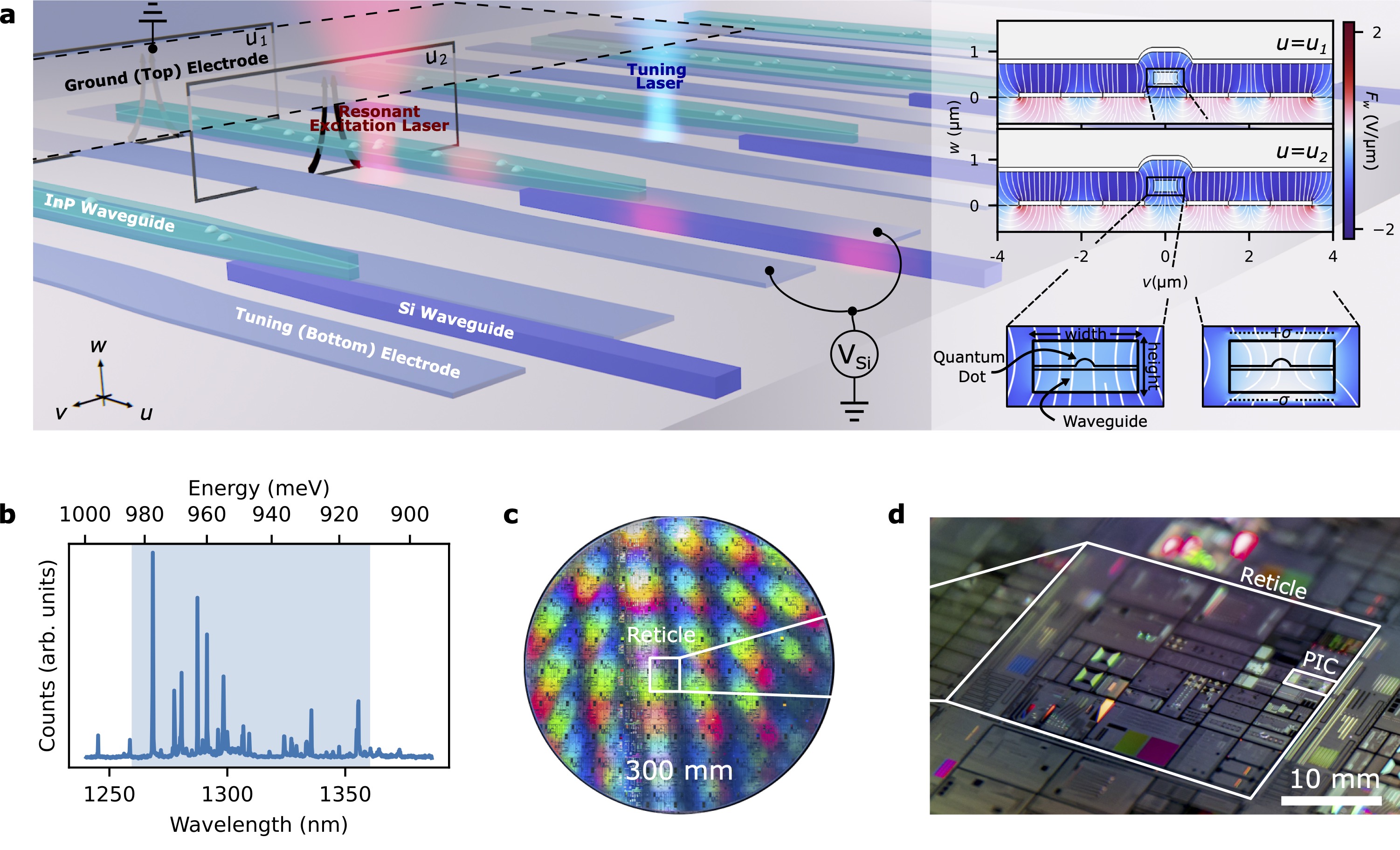}
	\caption{\textbf{Hybrid Integration Architecture.} {\bf a,} Schematics illustrating our approach to integrating tunable single photon emitters embedded in a transfer-printed InP chiplet coupled to a Si PIC waveguide. Maintaining the alignment between the waveguides of the InP chiplet and the Si PIC enables single photon adiabatic coupling in the resulting hybrid structure. We resonantly excite the emitters by focusing a laser beam onto the chiplet's emitters. Likewise, focusing a laser beam with an energy above the bandgap of the chiplet allows non-volatile tuning of the emitters. Specifically, applying a voltage, $V_\text{Si}$, between a tuning and a ground electrode placed beneath and over the chiplet, respectively, results in an electric field, represented as black lines, across the InP waveguide. This field redistributes carriers excited by the tuning laser in the chiplet, thereby screening the electric field applied to the QDs and locally altering their emission spectra. Inset: finite element simulations of the electric field, white lines overlaid on the component $F_w$ normal to the PIC surface, inside the InP waveguide due to carrier redistribution modeled as charged planes beneath and above the chiplet. QDs can experience various degrees of screening based on their location, thus resulting in non-volatile shifts in their emission frequencies. To illustrate, the emitter located at $u=u_1$ experiences no screening whereas the one at $u=u_2$ experiences screening due to surface charge densities $\pm\sigma$ enclosing the waveguide. {\bf b,} Representative photoluminescence spectrum, where the blue shading denotes the O-band, collected from photons that have been routed through the PIC. {\bf c,} Photograph of a 300~mm wafer on which our PIC was fabricated. {\bf d,} Closer view of the wafer from {\bf c} showing the location of our chip.}
	\label{fig:fig1}
\end{figure*}

\begin{figure*}[t]
	\centering
	\includegraphics[width=\linewidth]{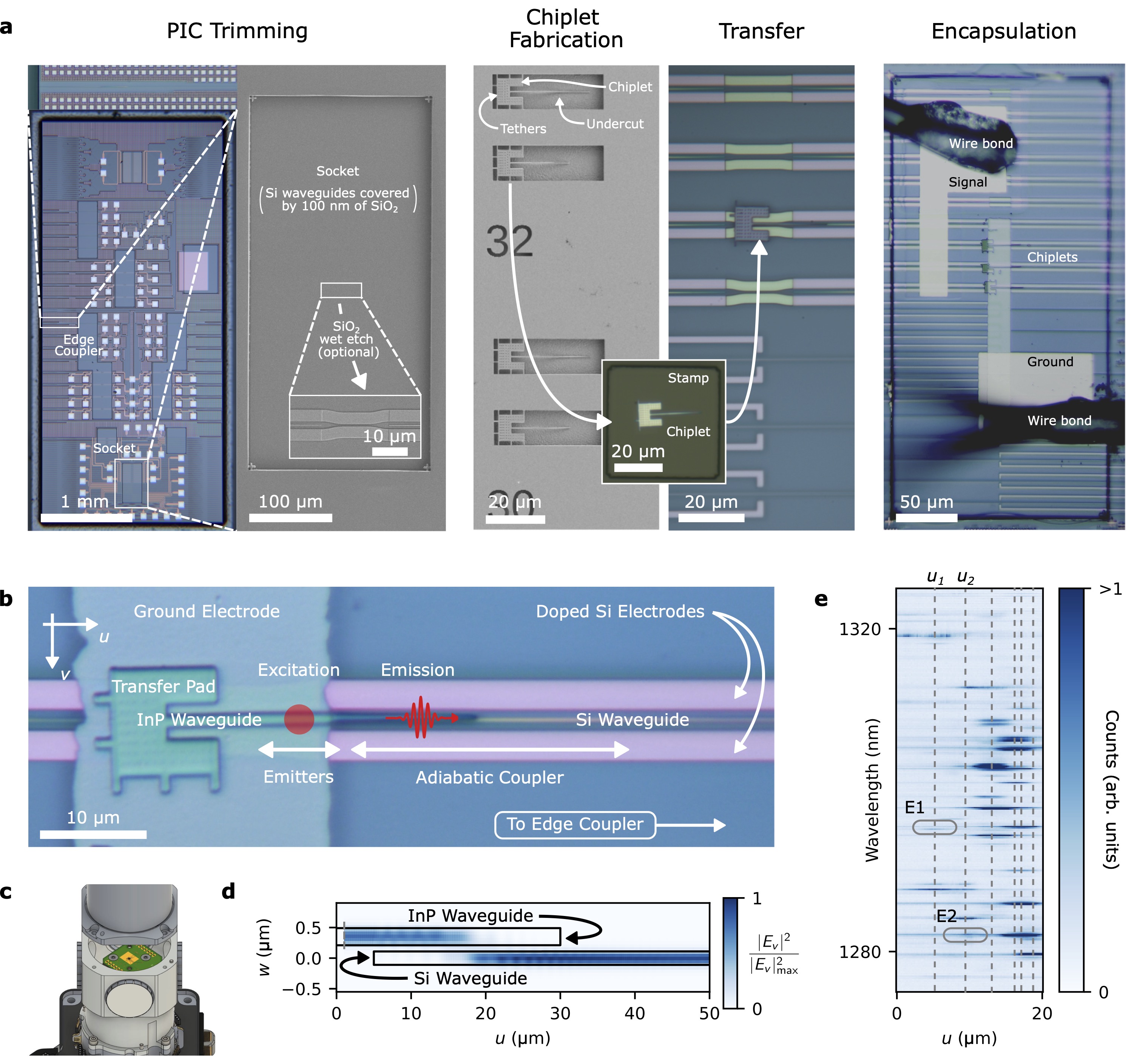}
	\caption{\textbf{Hybrid PIC Assembly.} {\bf a,} Images of the hybrid PIC at various stages of its assembly along with some of the components used in the process. {\bf b,} Optical micrograph of the transfered microchiplet surrounded by the components tuning its emitters and routing single photons to the Si PIC. {\bf c,} Schematics of the 5~K cryostat enclosing the PIC during its operation. {\bf d,} Finite difference time domain simulation of the InP waveguide quasi TE mode coupling from the chiplet to the PIC, where $E_v$ denotes the out-of-plane component of the electric field. \textbf{e,} PL spectra acquired while exciting emitters located at various positions along the chiplet's nanobeam.}
	\label{fig:fig2}
\end{figure*}

\begin{figure*}[t]
	\centering
	\includegraphics[width=\linewidth]{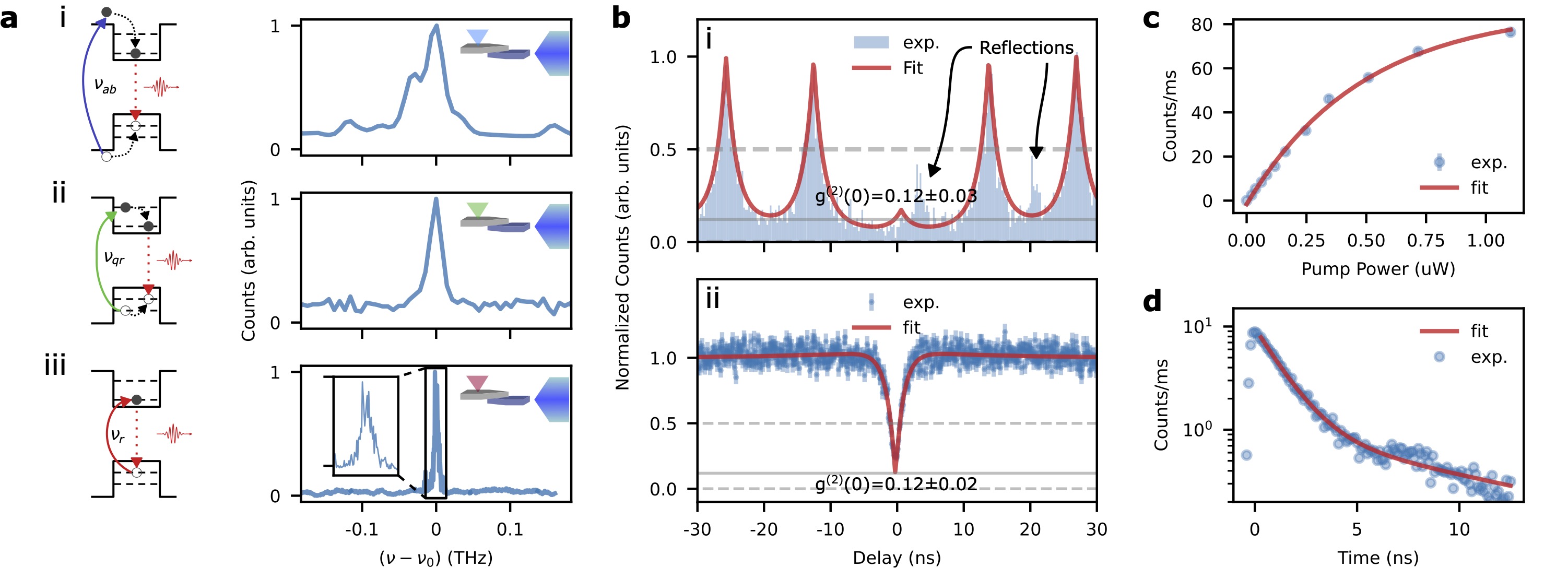}
	\caption{\textbf{Photon Statistics.} {\bf a,} Emission spectra of emitter 1 (E1) under (i) above-band, (ii) quasi-resonant, and (iii) resonant excitation. All spectra are horizontally shifted by the emission frequency of the emitter, $\nu_0$. {\bf b,} Autocorrelation measurement under (i) triggered and (ii) continuous wave excitation, both of which with a fitted $g^{(2)}(0)$ value of $0.12$. {\bf c,} Saturation curve of the emitter under pulsed above-band excitation. {\bf d} Radiative decay time trace of the emitter under pulsed above-band excitation.}	
    \label{fig:fig3}
\end{figure*}

\textbf{Hybrid architecture --} Figure~\ref{fig:fig1}a illustrates our hybrid architecture, where we integrate the SOI PIC with InP waveguide chiplets containing a central layer of InAs/InP QDs. As shown in the inset, the segmented top and bottom electrodes enable the application of electric fields primarily perpendicular to the QD layer to tune exciton emission via the quantum confined Stark effect~\cite{Miller:84i, Miller:84ii}. We realize these electrodes with segmented bottom electrodes in $p$-doped Si and a top-deposited ground electrode. 

\textbf{Non-volatile memory Stark shift control --} Spectrally tuning individual QDs along the waveguide direction, $u$, would require complex control electronics while also placing extreme demands on this bottom electrode layer in terms of spatial resolution and wiring. Therefore, we implemented an entirely different approach to Stark shift control at individual QD locations:  gate-controlled charge trapping of photogenerated carriers. Modeled by surface charges of $+\sigma$ ($-\sigma$) on the top (bottom) of the InP waveguide, these act as non-volatile floating gates that modulate the electric field across the QD layer, as seen by comparing the waveguide cross-sections at $u=u_1$ and $u=u_2$ in the inset of Fig.~\ref{fig:fig1}a. 

\textbf{Global telecom QD properties --} We use confocal excitation to program the memory and pump the QDs. QD emission into an InP waveguide chiplet thereafter couples via adiabatically tapered inverse couplers to the SOI PIC's waveguide layer for on-chip routing or edge-coupling into silica fibers for off-chip routing. Figure~\ref{fig:fig1}b plots a representative photoluminescence (PL) spectrum collected through the edge-couplers at a base temperature of 5~K and a pump laser wavelength $\lambda_p=780$~nm at 0.5~\textmu W focused to a $0.92$~\textmu m spot-size. The spectrum shows QD emission in multiple, spectrally distinct lines in the telecom O-band; their density is consistent with $\sim$ 10 QDs~\textmu m$^{-2}$.

\textbf{Foundry PIC design \& fabrication --} We fabricated the PICs in the AIM Photonics' foundry. Using 193~nm deep-ultraviolet water-immersion lithography, this fabrication process enables 300~mm wafer-scale production of SOI PICs with multiple metal and dielectric layers along with available $p$- and $n$-type doping. The PIC used an O-band specific process development kit element for broadband optical edge couplers featuring loss below 3 dB over a 1260-1360~nm wavelength range~\cite{Timurdogan:19} and other components custom designed for O-band operation. Figure~\ref{fig:fig1}c shows a wafer before dicing. Figure~\ref{fig:fig1}d highlights the PIC in this study, which occupies a $2\times 5$~mm$^{2}$ block of the reticle.

\subsection{Hybrid Integration}

\textbf{Postprocessing and chiplet integration --} At this stage, all waveguides are still covered in oxide and metal layers, as required by the foundry. To integrate quantum devices, it was necessary to develop a series of post-processing steps suited for a university cleanroom. Starting from the foundry-provided PIC, Fig.~\ref{fig:fig2}a summarizes the key stages of postprocessing which begins with the fabrication of `quantum sockets' from the PIC surface to the Si waveguides by optical lithography and wet etching. We leave a 100~nm oxide layer over the waveguides to ensure good mechanical adhesion and optical coupling between the chiplet and the PIC. In parallel, we fabricate suspended InP microchiplets with embedded InAs/InP QDs~\cite{Lee:20} by electron beam lithography and a combination of dry and wet etching. We then proceed  with transfer printing microchiplets from the parent InP chip into the PIC quantum sockets. This pick-and-place procedure uses a PDMS microstamp (50 \textmu m x 50 \textmu m) tracked under a microscope, achieving a chiplet placement accuracy of $44 $~nm (see SI). Finally, in the `encapsulation' step, we deposit a 475~nm thick PECVD oxide spacer layer and pattern a 20~nm Cr top electrode onto the quantum socket~\cite{Aghaeimeibodi:19}. 

Figure~\ref{fig:fig2}b shows an integrated microchiplet at the end of the fabrication process. Here, the ground electrode covers the section of the nanobeam before the chiplet's adiabatic taper. Our experiments target emitters between the ground and doped silicon electrodes as they strongly overlap with the electric field of this capacitive structure. After wirebonding to a cryo-compatible printed circuit board, we mount the PIC in a 5~K cryostat as illustrated in Fig.~\ref{fig:fig2}c. With a confocal microscope setup, we excite the QDs embedded in the InP chiplet through a window in the cryostat exposing the top of the device. 

\textbf{Chiplet to SOI waveguide coupling --} The  finite difference time domain simulation in Fig.~\ref{fig:fig2}d indicate that our optimized adiabatic couplers can reach a photon transfer efficiency of up to 99.5\%. Fiber-to-fiber measurements through a 2~mm silicon waveguide with a gap bridged by a double-sided InP taper reveal that this adiabatic transfer can reach efficiencies of $86 \pm 1 \%$ in practice. We extract this value by comparing transmission through the structure with transmission through a straight silicon waveguide. The latter measurement indicates a total transmission of $25 \pm 5 \%$ while using 5~\textmu m mode field diameter lensed fibers, indicating fiber-to-waveguide facet transmission better than 50\% enabled by the multiple dielectric layers in this SOI PIC platform. 

\textbf{Low-temperature QD-PIC spectroscopy --} Figure~\ref{fig:fig2}e presents a density plot of PL spectra under the same conditions as Fig.~\ref{fig:fig1}b, acquired while sweeping the excitation laser spot along  the waveguide axis $u$ while collecting out of the PIC facet into a single mode fiber. This plot shows a number of distinct QD emission lines over a spatial extent of $\sim 1.5$~\textmu m, which is consistent with the laser spot size. We will now consider detailed spectroscopy on emitters E1 and E2 at locations $u_1$ and $u_2$.

\subsection{Photon Statistics}

\textbf{Photon statistics measurements -} To demonstrate the versatility and performance of our system, we performed low-temperature photon correlation measurements on the PIC output facet under the three laser excitation regimes indicated in Fig.~\ref{fig:fig3}a: (i) above-band pumping, (ii) quasi-resonant excitation of a high-order excited state, and (iii) resonant excitation. The PL spectrum narrows to a linewidth of $15.5 \pm 0.6$~GHz when going from above-band to quasi-resonant excitation; this is near the 10~GHz resolution limit of the spectrometer that we use to measure (i) and (ii). Figure~\ref{fig:fig3}a(iii) shows the photoluminescence excitation obtained by scanning an external cavity diode 200~kHz linewidth laser across the emitter at a laser intensity of $\sim$ 10~\textmu W/\textmu m$^2$ while measuring the PIC output with superconducting nanowire single photon detectors (SNSPDs). As seen from the inset, we estimate a linewidth of $5.8\pm 0.2$~GHz. We found it unnecessary to use off-chip filtering under any excitation conditions during emission spectrum measurements, including the resonant case.  

\begin{figure*}[t]
	\centering
	\includegraphics[width=\linewidth]{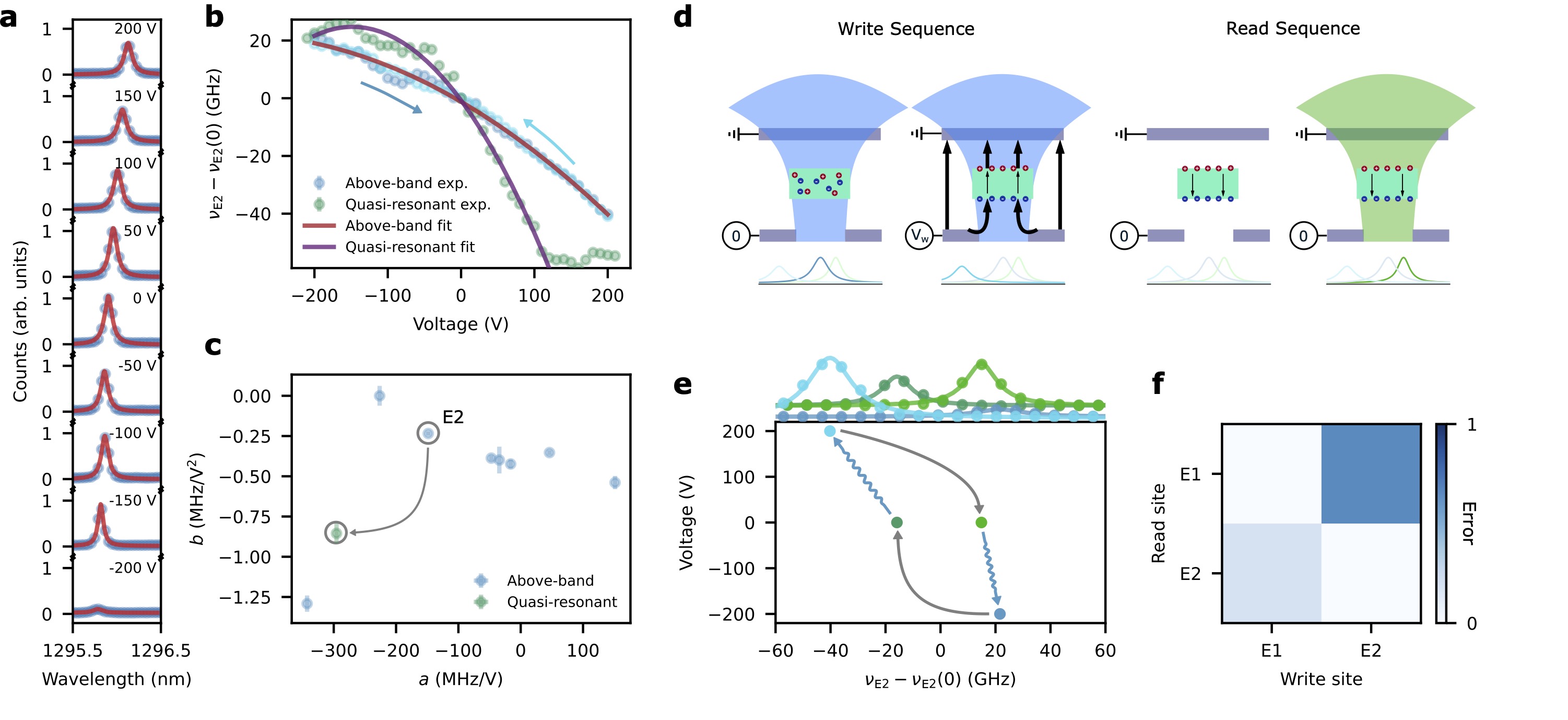}
	\caption{\textbf{Tuning.} {\bf a,} Emission spectrum and corresponding Lorentzian fits of a quantum dot under above-band illumination while voltages within $\pm 200$~V are applied to the device. {\bf b,} Center wavelength of the Lorentzian fits shown in \textbf{a} in addition to data for other applied voltages and while the emitter is quasi-resonantly excited. {\bf c,} Best quadratic fit parameters to the shift experienced by eight different emitters extracted by fitting data of the type shown in \textbf{b}. {\bf d,} Sequence taken to alter the internal electric field in the III-V chiplet. Above-band illumination first excites carriers in the chiplet. Applying a voltage to the device then redistributes the carriers in the device. Turning off the illumination and the voltage subsequently immobilizes the charges which apply an internal field  to the quantum dots. Emission variations under quasi-resonant excitation in turn provide information regarding this internal field. {\bf e,} Emission wavelength of the quantum dot during the steps shown in {\bf d} over two sequences where voltages of $\pm 200$~V were applied during above-band illumination. {\bf f,} Residual normalized electric field experienced by a quantum emitter while tuning the electric field of another with the sequence shown in \textbf{d}.}	
        \label{fig:fig4}
\end{figure*}

Next, we examine the purity of the excited photons with second order autocorrelation measurements using an off-chip Hanbury Brown-Twiss apparatus consisting of a 50:50 fiber splitter and SNSPDs. We first triggered above-band excitation with a pulsed 776~nm fiber-coupled laser with a 80~MHz repetition rate and a 120~fs pulse width. Figure~\ref{fig:fig3}b provides the resulting autocorrelation trace of the emitted photons after going through a 0.1~nm narrow-band tunable fiber filter that we use to remove residual counts from nearby emitters also excited by above-band pumping. In spite of spurious peaks resulting from imperfections in the exciting pulses (see SI), a fit of the data reveals clear anti-bunching at $g^{(2)}(0)=0.12 \pm 0.03$. Figure~\ref{fig:fig3}b also provides the result of similar measurements using a continuous wave tunable O-band laser resonantly exciting the emitter. The resonance fluorescence gives $g^{(2)}(0)=0.12 \pm 0.02$. This measurement did not require any filtering due to good pump rejection inherited from the perpendicular geometry of our excitation and collection optics~\cite{Muller:07, Huber:20}. Our normalized raw $g^{(2)}(0)$ is similar to the value reported in other works that focus on collecting filter-free resonance fluorescence from visible QDs~\cite{Huber:20}. Fitting our autocorrelation data to a model incorporating our SNSPDs' response time could imply a higher photon purity than what we obtain with our current fit. Additions to our measurement apparatus for improved photon purity through better pump rejection include polarization filtering~\cite{Wang:19}, temporal gating~\cite{Reithmaier:15}, or bi-chromatic pumping~\cite{He:19}.

Fitting the emitted photon flux against the pump intensity to $I(P)=I_\text{sat}(1-\exp(-P/P_\text{E1,sat}))$ indicates a saturation power of $P_\text{E1,sat}=0.32$~\textmu W. These measurements are carried out with above-band excitation using an objective lens with a numerical aperture of 0.55, hence a saturation intensity of $I_\text{E1,sat}=0.82$~\textmu W/\textmu m$^2$. We find that this power value varies from 0.14~\textmu W to 0.57~\textmu W based on measurements for five emitters (see SI). Figure~\ref{fig:fig3}c plots the saturation data and the corresponding fit for an emitter with a saturation power of $P_\text{sat}=0.49$~\textmu W. 

Figure~\ref{fig:fig3}d plots the time-resolved PL under pulsed above-band excitation, as measured on the SNSPDs. Fitting to a double-exponential decay $I(t)=a\exp(-t/\tau)+\tilde{a}\exp(-t/\tilde{\tau})$, where $a>\tilde{a}$,  gives $\tau=1.26$~ns, thereby suggesting a lifetime-limited linewidth of $\Delta \nu = 1/(2\pi\tau)=126~\text{MHz}$. We attribute $\tau$ and $\tilde \tau$ to bright excitons within well-defined QDs and adverse electron-hole recombination processes, respectively. Radiative decay traces for emitters with lower $\tau$ more prominently feature the influence of $\tilde \tau$ (see SI).

\subsection{QD spectral tuning by a non-volatile memory}

We quantify the tunability of our quantum socket by examining the PL spectra of emitter E2 under above-band excitation while increasing the voltage applied to the device. Figure~\ref{fig:fig4}a presents E2's PL spectrum under continuous-wave (cw) above-band excitation as a function of the electric potential difference $\Delta V=V_\text{Si}-V_\text{gnd}$ between the tuning electrodes and the ground-plane. As indicated in Fig.~\ref{fig:fig1}a, we expect this potential difference to apply a dominantly perpendicular electric field across the QDs. We observe a voltage-dependent shift in the emission line's center frequency $\nu_\text{E2}(\Delta V)$ at an approximately constant PL intensity from $\Delta V=-150$~V to $200$~V. We attribute the intensity drop at large negative potentials to local material variations, as each dot exhibits a different response (see SI). The PL linewidth and brightness is also approximately constant within roughly $|\Delta V| < 150$~V. $\nu(\Delta V)$ is also consistent up to the resolution of our spectrometer across multiple measurements, thus suggesting that our tuning mechanism does not induce any significant noise (see SI). 

Figure~\ref{fig:fig4}b plots the detuning $\nu_\text{E2}(\Delta V)-\nu_\text{E2}(0)$ obtained from Lorentzian fits to the spectra from Fig.~\ref{fig:fig4}a. Fitting these data to a quadratic model  $ \nu_\text{E2}(\Delta V) =  \nu_{0,\text{E2}} + a_\text{E2}\Delta V + b_\text{E2}\Delta V^2 $ gives a good agreement ($r^2=0.997$) for  $a_{\text{E2}}=-148 \pm 1$~MHz/V and $b_\text{E2}= -0.23 \pm 0.01 $~MHz/V$^2$. Figure~\ref{fig:fig4}c provides the fitted $a$ and $b$ for E2 and seven other emitters in the transferred chiplets. We observe an average $a=0 \pm 100$~MHz/V, whereas the $b$ parameter falls around $b= -0.3 \pm 0.2 $~MHz/V$^2$, indicating similar voltage sensitivities for our emitters.

The quadratic frequency shift with perpendicular electric field, $F_w$ agrees with the form of the quantum confined Stark effect: 
\begin{eqnarray}
    \label{eq:stark}
    h \Delta \nu &=& p F_w + \beta F_w^2 + \mathcal{O}(F_w^3), \\
    \label{eq:efield}
    F_w &=& \alpha(\Delta V/d) + F_0,\\
    \label{eq:p}
    p/h &=& a(d/\alpha) - 2 b (d/\alpha)^2 F_0, \\
    \label{eq:beta}
    \beta/h &=& b (d/\alpha)^2,
\end{eqnarray}
where $p$ and $\beta$ account for the dipole moment and the polarization of the dot, respectively. Here we introduce $d=2.44$~\textmu m as a characteristic electrode spacing weighted by the fringing in the electric field as seen in Fig.~\ref{fig:fig1}a; the free variables $\alpha$ and $F_0$ capture the reduction and offset of the electric field at the QD due to free carriers, respectively. Fitting this model with separately measured values of $p$ and $\beta$~\cite{Aghaeimeibodi:19} indicates that the strong electric field screening under above-band excitation ($\alpha_\text{ab}=0.017$) is improved under quasi-resonant excitation ($\alpha_\text{qr}=0.033$). Both conditions entail a similar offset field of $F_{0,\text{ab}}=59$~kV/cm and $F_{0,\text{qr}}=60$~kV/cm, respectively. 

To examine the role of charge distribution on the proposed screening, we proceed by applying a sequence of $\Delta V$ to the device under different illumination conditions. Figure~\ref{fig:fig4}d depicts the following steps of the sequence: write -- illuminating the chiplet with above-band light while ramping from $\Delta V=0$ to a ``write'' value $\Delta V= V_w$; read -- probing $\nu_\text{qr}(\Delta V = 0)$ under quasi-resonant excitation. Figure~\ref{fig:fig4}e plots this write-read sequence for E2. Starting from $(\Delta V,\nu_\text{E2}-\nu_\text{E2}(0))=(0~\text{V}, 15~\text{GHz})$ probed under quasi-resonant excitation, program the dot to ($0~\text{V},15~\text{GHz})\rightarrow (-200~\text{V},22~\text{GHz}) \rightarrow (0~\text{V},-16~\text{GHz}) $. Likewise, programming the dot to (200~\text{V},-40~\text{GHz}) under above-band illumination reverts the state of the dot back to $(0~\text{V},15~\text{GHz})$. This shows the ability to write $\nu_\text{E2}$ so that quasi-resonant excitation produces read values within 30~GHz. We attribute this memory effect to non-volatile charge redistribution near the QD.  

To characterize the cross-talk of this non-volatile memory, we perform an illumination sequence from Figs.~\ref{fig:fig4}d,e on one dot, e.g. dot $i$, followed by a measurement of the emission wavelength under quasi-resonant excitation. We then run an additional quasi-resonant measurement on a dot located $\sim 5$~\textmu m away from the first one, e.g. dot $j$. This procedure allows us to measure how changing the internal field applied to the one dot influences the internal field on the other. Specifically, for dot $j$, we obtain the frequency shift $\Delta \nu_{j}$ by letting $F_0 \rightarrow F_j + \delta F_{j,i}$ in Eq.~(\ref{eq:stark}), i.e.,
\begin{equation}
    h \Delta \nu_j = p_j (F_j + \delta F_{j,i}) + \beta_j (F_j + \delta F_{j,i})^2,
\end{equation}
where $F_{j}$ is the initial internal field on dot $j$ and $\delta F_{j,i}$ is the modification of this field following the above-band illumination of dot $i$. Partial exposure to the optical field illuminating dot $i$ may cause a cross-talk on dot $j$, i.e., $\delta F_{j,i} \neq 0$. In practice, an aberrated illumination beam may cause such cross-talk. Even if completely mitigated, the diffraction limit will still lower-bound $\delta F_{j,i}$. We quantify the error from this cross-talk by $|(\delta_{j,i} \, \delta F_{j,j} - \delta F_{j,i})/\delta F_{j,j}|$, where $\delta_{j,i}$ is the Kronecker $\delta$ function. As seen from the cross-talk matrix in Fig.~\ref{fig:fig4}f, we observe a maximum cross-talk of 0.65 in the off-diagonals for emitters separated by roughly 5~\textmu m.

\section{Discussion}

Future work should consider the following approaches to increasing the spectral tuning range of our device, narrowing the emitter linewidth, and leveraging other capabilites of advanced SOI PIC platforms. Probing the emission of QDs between the transfer and encapsulation steps may provide insight into how the processing steps affect the coherence of the emitted photons. Such measurements would lead to an improved post-processing flow able to narrow the emission linewidths of the QDs, leading to charge reductions and a path towards widely-tunable and lifetime-limited on-chip SPEs at telecom wavelengths. Furthermore, modifying the design for charging a tuning component without direct contact to the transferred dots could provide a localized, non-volatile, and large-scale tuning mechanism for $\mathcal{O}(N)$ QDs with $\mathcal{O}(1)$ electrodes without altering the coherence or the tuning sensitivity of emitters. Inverting error matrices similar to the one in Fig.~\ref{fig:fig4}f could provide a means to mitigate cross-talk while tuning $N$ such emitters, hence enabling applications  requiring multiple resonant SPEs.

The large-scale integration capabilities of silicon photonics make it ideally suited for hosting greater amounts of SPEs. Increasing the number of integrated chiplets would require a scalable transfer method, drawing on existing transfer printing techniques~\cite{Meitl:06} used for large-scale hybrid integration of other light sources such as colloidal QDs and lasers~\cite{Kim:11, Justice:12}. Devices with large numbers of emitters could lead towards more sophisticated on-chip many-photon quantum systems that may also benefit from active components enabled by the platform such as RF electronics or cryogenics-compatible optical modulators~\cite{Chakraborty:20}. Integrating other components such as on-chip SNSPDs~\cite{Ferrari:18,Gyger:21} could also enable more complex types of on-chip photonic computations, including more complex quantum information processing tasks~\cite{Rudolph:17} and machine learning~\cite{Steinbrecher:19, Lopez:21}. 

A combination of QD positioning methods used during chiplet fabrication~\cite{Liu:18, Schnauber:19} with large-scale schemes for optically probing nanophotonic structures~\cite{Panuski:22} could efficiently prepare large numbers of QD chiplets for transfer. Preliminary screening of optical properties of emitters and improving their tunability beyond 100~GHz could lead towards on-chip sources spanning telecommunication windows.

In summary, we introduced a platform for integrating tunable single-photon emitters on large-scale PICs. 
By hybrid integration of InAs/InP chiplets with a state-of-the-art SOI PIC, we have demonstrated scalable emission wavelength programmability with individual emitter resolution.
The excellent pump rejection of the PIC enables the observation of QD resonance fluorescence without additional filtering. The localized non-volatile spectral tuning opens the door to QD wavelength tuning at the scale of the number of resolvable laser spots, which can be in the millions with conventional microscopes. 
The combined control of photonic and atomic systems on a scalable platform enables a new generation of programmable quantum information processors manufactured in leading semiconductor foundries that can leverage many-body quantum systems.

\noindent \textbf{Acknowledgements}
H.L. and D.E. acknowledge support from the NSF (1936314QII-TAQS) and AFOSR program FA9550-16-1-0391. H. L. acknowledges the support of the Natural Sciences and Engineering Research Council of Canada (NSERC), the National Science Foundation (NSF, Award no. ECCS-1933556), and of the QISE-NET program of the NSF (NSF award DMR-1747426).
C.E-H. acknowledges funding from the European Union’s Horizon 2020 research and innovation program under the Marie Sklodowska-Curie grant agreement No.896401.
M.A.B., S.H, C.-M.L., and E.W. acknowledge financial support from the National Science Foundation (grants \#OMA1936314, \#OMA2120757, \#PHYS1915375, and \#ECCS1933546), AFOSR (grant \#FA23862014072), and the Maryland-ARL Quantum Partnership (W911NF1920181).
The authors acknowledge James M. Daley for assisting with the sample's fabrication along with Marco Colangelo, Alessandro Buzzi, and Camille Papon for useful discussions. The authors acknowledge MIT.nano's facilities where additional fabrication work was performed. The simulations were performed using RSoft Photonic Device Tools distributed through Synospys' University Software Program. The authors acknowledge X-Celeprint for distributing the PDMS stamps used in the assembly of the hybrid chip.


\begin{thebibliography}{0}%
\makeatletter
\providecommand \@ifxundefined [1]{%
 \@ifx{#1\undefined}
}%
\providecommand \@ifnum [1]{%
 \ifnum #1\expandafter \@firstoftwo
 \else \expandafter \@secondoftwo
 \fi
}%
\providecommand \@ifx [1]{%
 \ifx #1\expandafter \@firstoftwo
 \else \expandafter \@secondoftwo
 \fi
}%
\providecommand \natexlab [1]{#1}%
\providecommand \enquote  [1]{``#1''}%
\providecommand \bibnamefont  [1]{#1}%
\providecommand \bibfnamefont [1]{#1}%
\providecommand \citenamefont [1]{#1}%
\providecommand \href@noop [0]{\@secondoftwo}%
\providecommand \href [0]{\begingroup \@sanitize@url \@href}%
\providecommand \@href[1]{\@@startlink{#1}\@@href}%
\providecommand \@@href[1]{\endgroup#1\@@endlink}%
\providecommand \@sanitize@url [0]{\catcode `\\12\catcode `\$12\catcode
  `\&12\catcode `\#12\catcode `\^12\catcode `\_12\catcode `\%12\relax}%
\providecommand \@@startlink[1]{}%
\providecommand \@@endlink[0]{}%
\providecommand \url  [0]{\begingroup\@sanitize@url \@url }%
\providecommand \@url [1]{\endgroup\@href {#1}{\urlprefix }}%
\providecommand \urlprefix  [0]{URL }%
\providecommand \Eprint [0]{\href }%
\providecommand \doibase [0]{http://dx.doi.org/}%
\providecommand \selectlanguage [0]{\@gobble}%
\providecommand \bibinfo  [0]{\@secondoftwo}%
\providecommand \bibfield  [0]{\@secondoftwo}%
\providecommand \translation [1]{[#1]}%
\providecommand \BibitemOpen [0]{}%
\providecommand \bibitemStop [0]{}%
\providecommand \bibitemNoStop [0]{.\EOS\space}%
\providecommand \EOS [0]{\spacefactor3000\relax}%
\providecommand \BibitemShut  [1]{\csname bibitem#1\endcsname}%
\let\auto@bib@innerbib\@empty
\end{thebibliography}%


\begin{thebibliography}{63}%
\makeatletter
\providecommand \@ifxundefined [1]{%
 \@ifx{#1\undefined}
}%
\providecommand \@ifnum [1]{%
 \ifnum #1\expandafter \@firstoftwo
 \else \expandafter \@secondoftwo
 \fi
}%
\providecommand \@ifx [1]{%
 \ifx #1\expandafter \@firstoftwo
 \else \expandafter \@secondoftwo
 \fi
}%
\providecommand \natexlab [1]{#1}%
\providecommand \enquote  [1]{``#1''}%
\providecommand \bibnamefont  [1]{#1}%
\providecommand \bibfnamefont [1]{#1}%
\providecommand \citenamefont [1]{#1}%
\providecommand \href@noop [0]{\@secondoftwo}%
\providecommand \href [0]{\begingroup \@sanitize@url \@href}%
\providecommand \@href[1]{\@@startlink{#1}\@@href}%
\providecommand \@@href[1]{\endgroup#1\@@endlink}%
\providecommand \@sanitize@url [0]{\catcode `\\12\catcode `\$12\catcode
  `\&12\catcode `\#12\catcode `\^12\catcode `\_12\catcode `\%12\relax}%
\providecommand \@@startlink[1]{}%
\providecommand \@@endlink[0]{}%
\providecommand \url  [0]{\begingroup\@sanitize@url \@url }%
\providecommand \@url [1]{\endgroup\@href {#1}{\urlprefix }}%
\providecommand \urlprefix  [0]{URL }%
\providecommand \Eprint [0]{\href }%
\providecommand \doibase [0]{http://dx.doi.org/}%
\providecommand \selectlanguage [0]{\@gobble}%
\providecommand \bibinfo  [0]{\@secondoftwo}%
\providecommand \bibfield  [0]{\@secondoftwo}%
\providecommand \translation [1]{[#1]}%
\providecommand \BibitemOpen [0]{}%
\providecommand \bibitemStop [0]{}%
\providecommand \bibitemNoStop [0]{.\EOS\space}%
\providecommand \EOS [0]{\spacefactor3000\relax}%
\providecommand \BibitemShut  [1]{\csname bibitem#1\endcsname}%
\let\auto@bib@innerbib\@empty
\bibitem [{\citenamefont {He}\ \emph {et~al.}(2013)\citenamefont {He},
  \citenamefont {He}, \citenamefont {Wei}, \citenamefont {Wu}, \citenamefont
  {Atat{\"u}re}, \citenamefont {Schneider}, \citenamefont {H{\"o}fling},
  \citenamefont {Kamp}, \citenamefont {Lu},\ and\ \citenamefont {Pan}}]{He:13}%
  \BibitemOpen
  \bibfield  {author} {\bibinfo {author} {\bibfnamefont {Y.-M.}\ \bibnamefont
  {He}}, \bibinfo {author} {\bibfnamefont {Y.}~\bibnamefont {He}}, \bibinfo
  {author} {\bibfnamefont {Y.-J.}\ \bibnamefont {Wei}}, \bibinfo {author}
  {\bibfnamefont {D.}~\bibnamefont {Wu}}, \bibinfo {author} {\bibfnamefont
  {M.}~\bibnamefont {Atat{\"u}re}}, \bibinfo {author} {\bibfnamefont
  {C.}~\bibnamefont {Schneider}}, \bibinfo {author} {\bibfnamefont
  {S.}~\bibnamefont {H{\"o}fling}}, \bibinfo {author} {\bibfnamefont
  {M.}~\bibnamefont {Kamp}}, \bibinfo {author} {\bibfnamefont {C.-Y.}\
  \bibnamefont {Lu}}, \ and\ \bibinfo {author} {\bibfnamefont {J.-W.}\
  \bibnamefont {Pan}},\ }\href {\doibase 10.1038/nnano.2012.262} {\bibfield
  {journal} {\bibinfo  {journal} {Nat. Nanotechnol.}\ }\textbf {\bibinfo
  {volume} {8}},\ \bibinfo {pages} {213} (\bibinfo {year} {2013})}\BibitemShut
  {NoStop}%
\bibitem [{\citenamefont {Somaschi}\ \emph {et~al.}(2016)\citenamefont
  {Somaschi}, \citenamefont {Giesz}, \citenamefont {De~Santis}, \citenamefont
  {Loredo}, \citenamefont {Almeida}, \citenamefont {Hornecker}, \citenamefont
  {Portalupi}, \citenamefont {Grange}, \citenamefont {Ant{\'o}n}, \citenamefont
  {Demory}, \citenamefont {G{\'o}mez}, \citenamefont {Sagnes}, \citenamefont
  {Lanzillotti-Kimura}, \citenamefont {Lema{\'\i}tre}, \citenamefont
  {Auffeves}, \citenamefont {White}, \citenamefont {Lanco},\ and\ \citenamefont
  {Senellart}}]{Somaschi:16}%
  \BibitemOpen
  \bibfield  {author} {\bibinfo {author} {\bibfnamefont {N.}~\bibnamefont
  {Somaschi}}, \bibinfo {author} {\bibfnamefont {V.}~\bibnamefont {Giesz}},
  \bibinfo {author} {\bibfnamefont {L.}~\bibnamefont {De~Santis}}, \bibinfo
  {author} {\bibfnamefont {J.~C.}\ \bibnamefont {Loredo}}, \bibinfo {author}
  {\bibfnamefont {M.~P.}\ \bibnamefont {Almeida}}, \bibinfo {author}
  {\bibfnamefont {G.}~\bibnamefont {Hornecker}}, \bibinfo {author}
  {\bibfnamefont {S.~L.}\ \bibnamefont {Portalupi}}, \bibinfo {author}
  {\bibfnamefont {T.}~\bibnamefont {Grange}}, \bibinfo {author} {\bibfnamefont
  {C.}~\bibnamefont {Ant{\'o}n}}, \bibinfo {author} {\bibfnamefont
  {J.}~\bibnamefont {Demory}}, \bibinfo {author} {\bibfnamefont
  {C.}~\bibnamefont {G{\'o}mez}}, \bibinfo {author} {\bibfnamefont
  {I.}~\bibnamefont {Sagnes}}, \bibinfo {author} {\bibfnamefont {N.~D.}\
  \bibnamefont {Lanzillotti-Kimura}}, \bibinfo {author} {\bibfnamefont
  {A.}~\bibnamefont {Lema{\'\i}tre}}, \bibinfo {author} {\bibfnamefont
  {A.}~\bibnamefont {Auffeves}}, \bibinfo {author} {\bibfnamefont {A.~G.}\
  \bibnamefont {White}}, \bibinfo {author} {\bibfnamefont {L.}~\bibnamefont
  {Lanco}}, \ and\ \bibinfo {author} {\bibfnamefont {P.}~\bibnamefont
  {Senellart}},\ }\href {\doibase 10.1038/nphoton.2016.23} {\bibfield
  {journal} {\bibinfo  {journal} {Nat. Photon.}\ }\textbf {\bibinfo {volume}
  {10}},\ \bibinfo {pages} {340} (\bibinfo {year} {2016})}\BibitemShut
  {NoStop}%
\bibitem [{\citenamefont {Ding}\ \emph {et~al.}(2016)\citenamefont {Ding},
  \citenamefont {He}, \citenamefont {Duan}, \citenamefont {Gregersen},
  \citenamefont {Chen}, \citenamefont {Unsleber}, \citenamefont {Maier},
  \citenamefont {Schneider}, \citenamefont {Kamp}, \citenamefont {H\"ofling},
  \citenamefont {Lu},\ and\ \citenamefont {Pan}}]{Ding:16}%
  \BibitemOpen
  \bibfield  {author} {\bibinfo {author} {\bibfnamefont {X.}~\bibnamefont
  {Ding}}, \bibinfo {author} {\bibfnamefont {Y.}~\bibnamefont {He}}, \bibinfo
  {author} {\bibfnamefont {Z.-C.}\ \bibnamefont {Duan}}, \bibinfo {author}
  {\bibfnamefont {N.}~\bibnamefont {Gregersen}}, \bibinfo {author}
  {\bibfnamefont {M.-C.}\ \bibnamefont {Chen}}, \bibinfo {author}
  {\bibfnamefont {S.}~\bibnamefont {Unsleber}}, \bibinfo {author}
  {\bibfnamefont {S.}~\bibnamefont {Maier}}, \bibinfo {author} {\bibfnamefont
  {C.}~\bibnamefont {Schneider}}, \bibinfo {author} {\bibfnamefont
  {M.}~\bibnamefont {Kamp}}, \bibinfo {author} {\bibfnamefont {S.}~\bibnamefont
  {H\"ofling}}, \bibinfo {author} {\bibfnamefont {C.-Y.}\ \bibnamefont {Lu}}, \
  and\ \bibinfo {author} {\bibfnamefont {J.-W.}\ \bibnamefont {Pan}},\ }\href
  {\doibase 10.1103/PhysRevLett.116.020401} {\bibfield  {journal} {\bibinfo
  {journal} {Phys. Rev. Lett.}\ }\textbf {\bibinfo {volume} {116}},\ \bibinfo
  {pages} {020401} (\bibinfo {year} {2016})}\BibitemShut {NoStop}%
\bibitem [{\citenamefont {Wang}\ \emph {et~al.}(2019)\citenamefont {Wang},
  \citenamefont {He}, \citenamefont {Chung}, \citenamefont {Hu}, \citenamefont
  {Yu}, \citenamefont {Chen}, \citenamefont {Ding}, \citenamefont {Chen},
  \citenamefont {Qin}, \citenamefont {Yang}, \citenamefont {Liu}, \citenamefont
  {Duan}, \citenamefont {Li}, \citenamefont {Gerhardt}, \citenamefont
  {Winkler}, \citenamefont {Jurkat}, \citenamefont {Wang}, \citenamefont
  {Gregersen}, \citenamefont {Huo}, \citenamefont {Dai}, \citenamefont {Yu},
  \citenamefont {H{\"o}fling}, \citenamefont {Lu},\ and\ \citenamefont
  {Pan}}]{Wang:19}%
  \BibitemOpen
  \bibfield  {author} {\bibinfo {author} {\bibfnamefont {H.}~\bibnamefont
  {Wang}}, \bibinfo {author} {\bibfnamefont {Y.-M.}\ \bibnamefont {He}},
  \bibinfo {author} {\bibfnamefont {T.~H.}\ \bibnamefont {Chung}}, \bibinfo
  {author} {\bibfnamefont {H.}~\bibnamefont {Hu}}, \bibinfo {author}
  {\bibfnamefont {Y.}~\bibnamefont {Yu}}, \bibinfo {author} {\bibfnamefont
  {S.}~\bibnamefont {Chen}}, \bibinfo {author} {\bibfnamefont {X.}~\bibnamefont
  {Ding}}, \bibinfo {author} {\bibfnamefont {M.~C.}\ \bibnamefont {Chen}},
  \bibinfo {author} {\bibfnamefont {J.}~\bibnamefont {Qin}}, \bibinfo {author}
  {\bibfnamefont {X.}~\bibnamefont {Yang}}, \bibinfo {author} {\bibfnamefont
  {R.-Z.}\ \bibnamefont {Liu}}, \bibinfo {author} {\bibfnamefont {Z.~C.}\
  \bibnamefont {Duan}}, \bibinfo {author} {\bibfnamefont {J.~P.}\ \bibnamefont
  {Li}}, \bibinfo {author} {\bibfnamefont {S.}~\bibnamefont {Gerhardt}},
  \bibinfo {author} {\bibfnamefont {K.}~\bibnamefont {Winkler}}, \bibinfo
  {author} {\bibfnamefont {J.}~\bibnamefont {Jurkat}}, \bibinfo {author}
  {\bibfnamefont {L.-J.}\ \bibnamefont {Wang}}, \bibinfo {author}
  {\bibfnamefont {N.}~\bibnamefont {Gregersen}}, \bibinfo {author}
  {\bibfnamefont {Y.-H.}\ \bibnamefont {Huo}}, \bibinfo {author} {\bibfnamefont
  {Q.}~\bibnamefont {Dai}}, \bibinfo {author} {\bibfnamefont {S.}~\bibnamefont
  {Yu}}, \bibinfo {author} {\bibfnamefont {S.}~\bibnamefont {H{\"o}fling}},
  \bibinfo {author} {\bibfnamefont {C.-Y.}\ \bibnamefont {Lu}}, \ and\ \bibinfo
  {author} {\bibfnamefont {J.-W.}\ \bibnamefont {Pan}},\ }\href {\doibase
  10.1038/s41566-019-0494-3} {\bibfield  {journal} {\bibinfo  {journal} {Nat.
  Photon.}\ }\textbf {\bibinfo {volume} {13}},\ \bibinfo {pages} {770}
  (\bibinfo {year} {2019})}\BibitemShut {NoStop}%
\bibitem [{\citenamefont {Senellart}\ \emph {et~al.}(2017)\citenamefont
  {Senellart}, \citenamefont {Solomon},\ and\ \citenamefont
  {White}}]{Senellart:17}%
  \BibitemOpen
  \bibfield  {author} {\bibinfo {author} {\bibfnamefont {P.}~\bibnamefont
  {Senellart}}, \bibinfo {author} {\bibfnamefont {G.}~\bibnamefont {Solomon}},
  \ and\ \bibinfo {author} {\bibfnamefont {A.}~\bibnamefont {White}},\ }\href
  {\doibase 10.1038/nnano.2017.218} {\bibfield  {journal} {\bibinfo  {journal}
  {Nat. Nanotechnol.}\ }\textbf {\bibinfo {volume} {12}},\ \bibinfo {pages}
  {1026} (\bibinfo {year} {2017})}\BibitemShut {NoStop}%
\bibitem [{\citenamefont {Tomm}\ \emph {et~al.}(2021)\citenamefont {Tomm},
  \citenamefont {Javadi}, \citenamefont {Antoniadis}, \citenamefont {Najer},
  \citenamefont {L{\"o}bl}, \citenamefont {Korsch}, \citenamefont {Schott},
  \citenamefont {Valentin}, \citenamefont {Wieck}, \citenamefont {Ludwig},\
  and\ \citenamefont {Warburton}}]{Tomm:21}%
  \BibitemOpen
  \bibfield  {author} {\bibinfo {author} {\bibfnamefont {N.}~\bibnamefont
  {Tomm}}, \bibinfo {author} {\bibfnamefont {A.}~\bibnamefont {Javadi}},
  \bibinfo {author} {\bibfnamefont {N.~O.}\ \bibnamefont {Antoniadis}},
  \bibinfo {author} {\bibfnamefont {D.}~\bibnamefont {Najer}}, \bibinfo
  {author} {\bibfnamefont {M.~C.}\ \bibnamefont {L{\"o}bl}}, \bibinfo {author}
  {\bibfnamefont {A.~R.}\ \bibnamefont {Korsch}}, \bibinfo {author}
  {\bibfnamefont {R.}~\bibnamefont {Schott}}, \bibinfo {author} {\bibfnamefont
  {S.~R.}\ \bibnamefont {Valentin}}, \bibinfo {author} {\bibfnamefont {A.~D.}\
  \bibnamefont {Wieck}}, \bibinfo {author} {\bibfnamefont {A.}~\bibnamefont
  {Ludwig}}, \ and\ \bibinfo {author} {\bibfnamefont {R.~J.}\ \bibnamefont
  {Warburton}},\ }\href {\doibase 10.1038/s41565-020-00831-x} {\bibfield
  {journal} {\bibinfo  {journal} {Nat. Nanotechnol.}\ }\textbf {\bibinfo
  {volume} {16}},\ \bibinfo {pages} {399} (\bibinfo {year} {2021})}\BibitemShut
  {NoStop}%
\bibitem [{\citenamefont {Thomas}\ \emph {et~al.}(2021)\citenamefont {Thomas},
  \citenamefont {Billard}, \citenamefont {Coste}, \citenamefont {Wein},
  \citenamefont {Priya}, \citenamefont {Ollivier}, \citenamefont {Krebs},
  \citenamefont {Taza\"{\i}rt}, \citenamefont {Harouri}, \citenamefont
  {Lemaitre}, \citenamefont {Sagnes}, \citenamefont {Anton}, \citenamefont
  {Lanco}, \citenamefont {Somaschi}, \citenamefont {Loredo},\ and\
  \citenamefont {Senellart}}]{Thomas:21}%
  \BibitemOpen
  \bibfield  {author} {\bibinfo {author} {\bibfnamefont {S.~E.}\ \bibnamefont
  {Thomas}}, \bibinfo {author} {\bibfnamefont {M.}~\bibnamefont {Billard}},
  \bibinfo {author} {\bibfnamefont {N.}~\bibnamefont {Coste}}, \bibinfo
  {author} {\bibfnamefont {S.~C.}\ \bibnamefont {Wein}}, \bibinfo {author}
  {\bibnamefont {Priya}}, \bibinfo {author} {\bibfnamefont {H.}~\bibnamefont
  {Ollivier}}, \bibinfo {author} {\bibfnamefont {O.}~\bibnamefont {Krebs}},
  \bibinfo {author} {\bibfnamefont {L.}~\bibnamefont {Taza\"{\i}rt}}, \bibinfo
  {author} {\bibfnamefont {A.}~\bibnamefont {Harouri}}, \bibinfo {author}
  {\bibfnamefont {A.}~\bibnamefont {Lemaitre}}, \bibinfo {author}
  {\bibfnamefont {I.}~\bibnamefont {Sagnes}}, \bibinfo {author} {\bibfnamefont
  {C.}~\bibnamefont {Anton}}, \bibinfo {author} {\bibfnamefont
  {L.}~\bibnamefont {Lanco}}, \bibinfo {author} {\bibfnamefont
  {N.}~\bibnamefont {Somaschi}}, \bibinfo {author} {\bibfnamefont {J.~C.}\
  \bibnamefont {Loredo}}, \ and\ \bibinfo {author} {\bibfnamefont
  {P.}~\bibnamefont {Senellart}},\ }\href {\doibase
  10.1103/PhysRevLett.126.233601} {\bibfield  {journal} {\bibinfo  {journal}
  {Phys. Rev. Lett.}\ }\textbf {\bibinfo {volume} {126}},\ \bibinfo {pages}
  {233601} (\bibinfo {year} {2021})}\BibitemShut {NoStop}%
\bibitem [{\citenamefont {Dousse}\ \emph {et~al.}(2010)\citenamefont {Dousse},
  \citenamefont {Suffczy{\'n}ski}, \citenamefont {Beveratos}, \citenamefont
  {Krebs}, \citenamefont {Lema{\^\i}tre}, \citenamefont {Sagnes}, \citenamefont
  {Bloch}, \citenamefont {Voisin},\ and\ \citenamefont
  {Senellart}}]{Dousse:10}%
  \BibitemOpen
  \bibfield  {author} {\bibinfo {author} {\bibfnamefont {A.}~\bibnamefont
  {Dousse}}, \bibinfo {author} {\bibfnamefont {J.}~\bibnamefont
  {Suffczy{\'n}ski}}, \bibinfo {author} {\bibfnamefont {A.}~\bibnamefont
  {Beveratos}}, \bibinfo {author} {\bibfnamefont {O.}~\bibnamefont {Krebs}},
  \bibinfo {author} {\bibfnamefont {A.}~\bibnamefont {Lema{\^\i}tre}}, \bibinfo
  {author} {\bibfnamefont {I.}~\bibnamefont {Sagnes}}, \bibinfo {author}
  {\bibfnamefont {J.}~\bibnamefont {Bloch}}, \bibinfo {author} {\bibfnamefont
  {P.}~\bibnamefont {Voisin}}, \ and\ \bibinfo {author} {\bibfnamefont
  {P.}~\bibnamefont {Senellart}},\ }\href {\doibase 10.1038/nature09148}
  {\bibfield  {journal} {\bibinfo  {journal} {Nature}\ }\textbf {\bibinfo
  {volume} {466}},\ \bibinfo {pages} {217} (\bibinfo {year}
  {2010})}\BibitemShut {NoStop}%
\bibitem [{\citenamefont {M{\"u}ller}\ \emph {et~al.}(2014)\citenamefont
  {M{\"u}ller}, \citenamefont {Bounouar}, \citenamefont {J{\"o}ns},
  \citenamefont {Gl{\"a}ssl},\ and\ \citenamefont {Michler}}]{Muller:14}%
  \BibitemOpen
  \bibfield  {author} {\bibinfo {author} {\bibfnamefont {M.}~\bibnamefont
  {M{\"u}ller}}, \bibinfo {author} {\bibfnamefont {S.}~\bibnamefont
  {Bounouar}}, \bibinfo {author} {\bibfnamefont {K.~D.}\ \bibnamefont
  {J{\"o}ns}}, \bibinfo {author} {\bibfnamefont {M.}~\bibnamefont
  {Gl{\"a}ssl}}, \ and\ \bibinfo {author} {\bibfnamefont {P.}~\bibnamefont
  {Michler}},\ }\href {\doibase 10.1038/nphoton.2013.377} {\bibfield  {journal}
  {\bibinfo  {journal} {Nat. Photon.}\ }\textbf {\bibinfo {volume} {8}},\
  \bibinfo {pages} {224} (\bibinfo {year} {2014})}\BibitemShut {NoStop}%
\bibitem [{\citenamefont {Liu}\ \emph {et~al.}(2019)\citenamefont {Liu},
  \citenamefont {Su}, \citenamefont {Wei}, \citenamefont {Yao}, \citenamefont
  {Silva}, \citenamefont {Yu}, \citenamefont {Iles-Smith}, \citenamefont
  {Srinivasan}, \citenamefont {Rastelli}, \citenamefont {Li},\ and\
  \citenamefont {Wang}}]{Liu:19}%
  \BibitemOpen
  \bibfield  {author} {\bibinfo {author} {\bibfnamefont {J.}~\bibnamefont
  {Liu}}, \bibinfo {author} {\bibfnamefont {R.}~\bibnamefont {Su}}, \bibinfo
  {author} {\bibfnamefont {Y.}~\bibnamefont {Wei}}, \bibinfo {author}
  {\bibfnamefont {B.}~\bibnamefont {Yao}}, \bibinfo {author} {\bibfnamefont
  {S.~F. C.~d.}\ \bibnamefont {Silva}}, \bibinfo {author} {\bibfnamefont
  {Y.}~\bibnamefont {Yu}}, \bibinfo {author} {\bibfnamefont {J.}~\bibnamefont
  {Iles-Smith}}, \bibinfo {author} {\bibfnamefont {K.}~\bibnamefont
  {Srinivasan}}, \bibinfo {author} {\bibfnamefont {A.}~\bibnamefont
  {Rastelli}}, \bibinfo {author} {\bibfnamefont {J.}~\bibnamefont {Li}}, \ and\
  \bibinfo {author} {\bibfnamefont {X.}~\bibnamefont {Wang}},\ }\href {\doibase
  10.1038/s41565-019-0435-9} {\bibfield  {journal} {\bibinfo  {journal} {Nat.
  Nanotechnol.}\ }\textbf {\bibinfo {volume} {14}},\ \bibinfo {pages} {586}
  (\bibinfo {year} {2019})}\BibitemShut {NoStop}%
\bibitem [{\citenamefont {Javadi}\ \emph {et~al.}(2015)\citenamefont {Javadi},
  \citenamefont {S{\"o}llner}, \citenamefont {Arcari}, \citenamefont {Hansen},
  \citenamefont {Midolo}, \citenamefont {Mahmoodian}, \citenamefont {Kir{\v
  s}ansk{\.e}}, \citenamefont {Pregnolato}, \citenamefont {Lee}, \citenamefont
  {Song}, \citenamefont {Stobbe},\ and\ \citenamefont {Lodahl}}]{Javadi:15}%
  \BibitemOpen
  \bibfield  {author} {\bibinfo {author} {\bibfnamefont {A.}~\bibnamefont
  {Javadi}}, \bibinfo {author} {\bibfnamefont {I.}~\bibnamefont {S{\"o}llner}},
  \bibinfo {author} {\bibfnamefont {M.}~\bibnamefont {Arcari}}, \bibinfo
  {author} {\bibfnamefont {S.~L.}\ \bibnamefont {Hansen}}, \bibinfo {author}
  {\bibfnamefont {L.}~\bibnamefont {Midolo}}, \bibinfo {author} {\bibfnamefont
  {S.}~\bibnamefont {Mahmoodian}}, \bibinfo {author} {\bibfnamefont
  {G.}~\bibnamefont {Kir{\v s}ansk{\.e}}}, \bibinfo {author} {\bibfnamefont
  {T.}~\bibnamefont {Pregnolato}}, \bibinfo {author} {\bibfnamefont {E.~H.}\
  \bibnamefont {Lee}}, \bibinfo {author} {\bibfnamefont {J.~D.}\ \bibnamefont
  {Song}}, \bibinfo {author} {\bibfnamefont {S.}~\bibnamefont {Stobbe}}, \ and\
  \bibinfo {author} {\bibfnamefont {P.}~\bibnamefont {Lodahl}},\ }\href
  {\doibase 10.1038/ncomms9655} {\bibfield  {journal} {\bibinfo  {journal}
  {Nat. Commun.}\ }\textbf {\bibinfo {volume} {6}},\ \bibinfo {pages} {8655}
  (\bibinfo {year} {2015})}\BibitemShut {NoStop}%
\bibitem [{\citenamefont {Le~Jeannic}\ \emph {et~al.}(2021)\citenamefont
  {Le~Jeannic}, \citenamefont {Ramos}, \citenamefont {Simonsen}, \citenamefont
  {Pregnolato}, \citenamefont {Liu}, \citenamefont {Schott}, \citenamefont
  {Wieck}, \citenamefont {Ludwig}, \citenamefont {Rotenberg}, \citenamefont
  {Garc\'{\i}a-Ripoll},\ and\ \citenamefont {Lodahl}}]{LeJeannic:21}%
  \BibitemOpen
  \bibfield  {author} {\bibinfo {author} {\bibfnamefont {H.}~\bibnamefont
  {Le~Jeannic}}, \bibinfo {author} {\bibfnamefont {T.}~\bibnamefont {Ramos}},
  \bibinfo {author} {\bibfnamefont {S.~F.}\ \bibnamefont {Simonsen}}, \bibinfo
  {author} {\bibfnamefont {T.}~\bibnamefont {Pregnolato}}, \bibinfo {author}
  {\bibfnamefont {Z.}~\bibnamefont {Liu}}, \bibinfo {author} {\bibfnamefont
  {R.}~\bibnamefont {Schott}}, \bibinfo {author} {\bibfnamefont {A.~D.}\
  \bibnamefont {Wieck}}, \bibinfo {author} {\bibfnamefont {A.}~\bibnamefont
  {Ludwig}}, \bibinfo {author} {\bibfnamefont {N.}~\bibnamefont {Rotenberg}},
  \bibinfo {author} {\bibfnamefont {J.~J.}\ \bibnamefont {Garc\'{\i}a-Ripoll}},
  \ and\ \bibinfo {author} {\bibfnamefont {P.}~\bibnamefont {Lodahl}},\ }\href
  {\doibase 10.1103/PhysRevLett.126.023603} {\bibfield  {journal} {\bibinfo
  {journal} {Phys. Rev. Lett.}\ }\textbf {\bibinfo {volume} {126}},\ \bibinfo
  {pages} {023603} (\bibinfo {year} {2021})}\BibitemShut {NoStop}%
\bibitem [{\citenamefont {Le~Jeannic}\ \emph {et~al.}(2022)\citenamefont
  {Le~Jeannic}, \citenamefont {Tiranov}, \citenamefont {Carolan}, \citenamefont
  {Ramos}, \citenamefont {Wang}, \citenamefont {Appel}, \citenamefont {Scholz},
  \citenamefont {Wieck}, \citenamefont {Ludwig}, \citenamefont {Rotenberg},
  \citenamefont {Midolo}, \citenamefont {Garc{\'\i}a-Ripoll}, \citenamefont
  {S{\o}rensen},\ and\ \citenamefont {Lodahl}}]{LeJeannic:22}%
  \BibitemOpen
  \bibfield  {author} {\bibinfo {author} {\bibfnamefont {H.}~\bibnamefont
  {Le~Jeannic}}, \bibinfo {author} {\bibfnamefont {A.}~\bibnamefont {Tiranov}},
  \bibinfo {author} {\bibfnamefont {J.}~\bibnamefont {Carolan}}, \bibinfo
  {author} {\bibfnamefont {T.}~\bibnamefont {Ramos}}, \bibinfo {author}
  {\bibfnamefont {Y.}~\bibnamefont {Wang}}, \bibinfo {author} {\bibfnamefont
  {M.~H.}\ \bibnamefont {Appel}}, \bibinfo {author} {\bibfnamefont
  {S.}~\bibnamefont {Scholz}}, \bibinfo {author} {\bibfnamefont {A.~D.}\
  \bibnamefont {Wieck}}, \bibinfo {author} {\bibfnamefont {A.}~\bibnamefont
  {Ludwig}}, \bibinfo {author} {\bibfnamefont {N.}~\bibnamefont {Rotenberg}},
  \bibinfo {author} {\bibfnamefont {L.}~\bibnamefont {Midolo}}, \bibinfo
  {author} {\bibfnamefont {J.~J.}\ \bibnamefont {Garc{\'\i}a-Ripoll}}, \bibinfo
  {author} {\bibfnamefont {A.~S.}\ \bibnamefont {S{\o}rensen}}, \ and\ \bibinfo
  {author} {\bibfnamefont {P.}~\bibnamefont {Lodahl}},\ }\href {\doibase
  10.1038/s41567-022-01720-x} {\bibfield  {journal} {\bibinfo  {journal} {Nat.
  Phys.}\ }\textbf {\bibinfo {volume} {18}},\ \bibinfo {pages} {1191} (\bibinfo
  {year} {2022})}\BibitemShut {NoStop}%
\bibitem [{\citenamefont {Lindner}\ and\ \citenamefont
  {Rudolph}(2009)}]{Lindner:09}%
  \BibitemOpen
  \bibfield  {author} {\bibinfo {author} {\bibfnamefont {N.~H.}\ \bibnamefont
  {Lindner}}\ and\ \bibinfo {author} {\bibfnamefont {T.}~\bibnamefont
  {Rudolph}},\ }\href {\doibase 10.1103/PhysRevLett.103.113602} {\bibfield
  {journal} {\bibinfo  {journal} {Phys. Rev. Lett.}\ }\textbf {\bibinfo
  {volume} {103}},\ \bibinfo {pages} {113602} (\bibinfo {year}
  {2009})}\BibitemShut {NoStop}%
\bibitem [{\citenamefont {Schwartz}\ \emph {et~al.}(2016)\citenamefont
  {Schwartz}, \citenamefont {Cogan}, \citenamefont {Schmidgall}, \citenamefont
  {Don}, \citenamefont {Gantz}, \citenamefont {Kenneth}, \citenamefont
  {Lindner},\ and\ \citenamefont {Gershoni}}]{Schwartz:16}%
  \BibitemOpen
  \bibfield  {author} {\bibinfo {author} {\bibfnamefont {I.}~\bibnamefont
  {Schwartz}}, \bibinfo {author} {\bibfnamefont {D.}~\bibnamefont {Cogan}},
  \bibinfo {author} {\bibfnamefont {E.~R.}\ \bibnamefont {Schmidgall}},
  \bibinfo {author} {\bibfnamefont {Y.}~\bibnamefont {Don}}, \bibinfo {author}
  {\bibfnamefont {L.}~\bibnamefont {Gantz}}, \bibinfo {author} {\bibfnamefont
  {O.}~\bibnamefont {Kenneth}}, \bibinfo {author} {\bibfnamefont {N.~H.}\
  \bibnamefont {Lindner}}, \ and\ \bibinfo {author} {\bibfnamefont
  {D.}~\bibnamefont {Gershoni}},\ }\href {\doibase 10.1126/science.aah4758}
  {\bibfield  {journal} {\bibinfo  {journal} {Science}\ }\textbf {\bibinfo
  {volume} {354}},\ \bibinfo {pages} {434} (\bibinfo {year}
  {2016})}\BibitemShut {NoStop}%
\bibitem [{\citenamefont {Istrati}\ \emph {et~al.}(2020)\citenamefont
  {Istrati}, \citenamefont {Pilnyak}, \citenamefont {Loredo}, \citenamefont
  {Ant{\'o}n}, \citenamefont {Somaschi}, \citenamefont {Hilaire}, \citenamefont
  {Ollivier}, \citenamefont {Esmann}, \citenamefont {Cohen}, \citenamefont
  {Vidro}, \citenamefont {Millet}, \citenamefont {Lema{\^\i}tre}, \citenamefont
  {Sagnes}, \citenamefont {Harouri}, \citenamefont {Lanco}, \citenamefont
  {Senellart},\ and\ \citenamefont {Eisenberg}}]{Istrati:20}%
  \BibitemOpen
  \bibfield  {author} {\bibinfo {author} {\bibfnamefont {D.}~\bibnamefont
  {Istrati}}, \bibinfo {author} {\bibfnamefont {Y.}~\bibnamefont {Pilnyak}},
  \bibinfo {author} {\bibfnamefont {J.~C.}\ \bibnamefont {Loredo}}, \bibinfo
  {author} {\bibfnamefont {C.}~\bibnamefont {Ant{\'o}n}}, \bibinfo {author}
  {\bibfnamefont {N.}~\bibnamefont {Somaschi}}, \bibinfo {author}
  {\bibfnamefont {P.}~\bibnamefont {Hilaire}}, \bibinfo {author} {\bibfnamefont
  {H.}~\bibnamefont {Ollivier}}, \bibinfo {author} {\bibfnamefont
  {M.}~\bibnamefont {Esmann}}, \bibinfo {author} {\bibfnamefont
  {L.}~\bibnamefont {Cohen}}, \bibinfo {author} {\bibfnamefont
  {L.}~\bibnamefont {Vidro}}, \bibinfo {author} {\bibfnamefont
  {C.}~\bibnamefont {Millet}}, \bibinfo {author} {\bibfnamefont
  {A.}~\bibnamefont {Lema{\^\i}tre}}, \bibinfo {author} {\bibfnamefont
  {I.}~\bibnamefont {Sagnes}}, \bibinfo {author} {\bibfnamefont
  {A.}~\bibnamefont {Harouri}}, \bibinfo {author} {\bibfnamefont
  {L.}~\bibnamefont {Lanco}}, \bibinfo {author} {\bibfnamefont
  {P.}~\bibnamefont {Senellart}}, \ and\ \bibinfo {author} {\bibfnamefont
  {H.~S.}\ \bibnamefont {Eisenberg}},\ }\href {\doibase
  10.1038/s41467-020-19341-4} {\bibfield  {journal} {\bibinfo  {journal} {Nat.
  Commun.}\ }\textbf {\bibinfo {volume} {11}},\ \bibinfo {pages} {5501}
  (\bibinfo {year} {2020})}\BibitemShut {NoStop}%
\bibitem [{\citenamefont {Cogan}\ \emph {et~al.}(2023)\citenamefont {Cogan},
  \citenamefont {Su}, \citenamefont {Kenneth},\ and\ \citenamefont
  {Gershoni}}]{Cogan:23}%
  \BibitemOpen
  \bibfield  {author} {\bibinfo {author} {\bibfnamefont {D.}~\bibnamefont
  {Cogan}}, \bibinfo {author} {\bibfnamefont {Z.-E.}\ \bibnamefont {Su}},
  \bibinfo {author} {\bibfnamefont {O.}~\bibnamefont {Kenneth}}, \ and\
  \bibinfo {author} {\bibfnamefont {D.}~\bibnamefont {Gershoni}},\ }\href
  {\doibase 10.1038/s41566-022-01152-2} {\bibfield  {journal} {\bibinfo
  {journal} {Nature Photonics}\ }\textbf {\bibinfo {volume} {17}},\ \bibinfo
  {pages} {324} (\bibinfo {year} {2023})}\BibitemShut {NoStop}%
\bibitem [{\citenamefont {Stockill}\ \emph {et~al.}(2016)\citenamefont
  {Stockill}, \citenamefont {Le~Gall}, \citenamefont {Matthiesen},
  \citenamefont {Huthmacher}, \citenamefont {Clarke}, \citenamefont {Hugues},\
  and\ \citenamefont {Atat{\"u}re}}]{Stockill:16}%
  \BibitemOpen
  \bibfield  {author} {\bibinfo {author} {\bibfnamefont {R.}~\bibnamefont
  {Stockill}}, \bibinfo {author} {\bibfnamefont {C.}~\bibnamefont {Le~Gall}},
  \bibinfo {author} {\bibfnamefont {C.}~\bibnamefont {Matthiesen}}, \bibinfo
  {author} {\bibfnamefont {L.}~\bibnamefont {Huthmacher}}, \bibinfo {author}
  {\bibfnamefont {E.}~\bibnamefont {Clarke}}, \bibinfo {author} {\bibfnamefont
  {M.}~\bibnamefont {Hugues}}, \ and\ \bibinfo {author} {\bibfnamefont
  {M.}~\bibnamefont {Atat{\"u}re}},\ }\href {\doibase 10.1038/ncomms12745}
  {\bibfield  {journal} {\bibinfo  {journal} {Nat. Commun.}\ }\textbf {\bibinfo
  {volume} {7}},\ \bibinfo {pages} {12745} (\bibinfo {year}
  {2016})}\BibitemShut {NoStop}%
\bibitem [{\citenamefont {Gangloff}\ \emph {et~al.}(2019)\citenamefont
  {Gangloff}, \citenamefont {Ethier-Majcher}, \citenamefont {Lang},
  \citenamefont {Denning}, \citenamefont {Bodey}, \citenamefont {Jackson},
  \citenamefont {Clarke}, \citenamefont {Hugues}, \citenamefont {Le~Gall},\
  and\ \citenamefont {Atat{\"u}re}}]{Gangloff:19}%
  \BibitemOpen
  \bibfield  {author} {\bibinfo {author} {\bibfnamefont {D.}~\bibnamefont
  {Gangloff}}, \bibinfo {author} {\bibfnamefont {G.}~\bibnamefont
  {Ethier-Majcher}}, \bibinfo {author} {\bibfnamefont {C.}~\bibnamefont
  {Lang}}, \bibinfo {author} {\bibfnamefont {E.}~\bibnamefont {Denning}},
  \bibinfo {author} {\bibfnamefont {J.}~\bibnamefont {Bodey}}, \bibinfo
  {author} {\bibfnamefont {D.}~\bibnamefont {Jackson}}, \bibinfo {author}
  {\bibfnamefont {E.}~\bibnamefont {Clarke}}, \bibinfo {author} {\bibfnamefont
  {M.}~\bibnamefont {Hugues}}, \bibinfo {author} {\bibfnamefont
  {C.}~\bibnamefont {Le~Gall}}, \ and\ \bibinfo {author} {\bibfnamefont
  {M.}~\bibnamefont {Atat{\"u}re}},\ }\href {\doibase 10.1126/science.aaw2906}
  {\bibfield  {journal} {\bibinfo  {journal} {Science}\ }\textbf {\bibinfo
  {volume} {364}},\ \bibinfo {pages} {62} (\bibinfo {year} {2019})}\BibitemShut
  {NoStop}%
\bibitem [{\citenamefont {Gangloff}\ \emph {et~al.}(2021)\citenamefont
  {Gangloff}, \citenamefont {Zaporski}, \citenamefont {Bodey}, \citenamefont
  {Bachorz}, \citenamefont {Jackson}, \citenamefont {{\'E}thier-Majcher},
  \citenamefont {Lang}, \citenamefont {Clarke}, \citenamefont {Hugues},
  \citenamefont {Le~Gall},\ and\ \citenamefont {Atat{\"u}re}}]{Gangloff:21}%
  \BibitemOpen
  \bibfield  {author} {\bibinfo {author} {\bibfnamefont {D.~A.}\ \bibnamefont
  {Gangloff}}, \bibinfo {author} {\bibfnamefont {L.}~\bibnamefont {Zaporski}},
  \bibinfo {author} {\bibfnamefont {J.~H.}\ \bibnamefont {Bodey}}, \bibinfo
  {author} {\bibfnamefont {C.}~\bibnamefont {Bachorz}}, \bibinfo {author}
  {\bibfnamefont {D.~M.}\ \bibnamefont {Jackson}}, \bibinfo {author}
  {\bibfnamefont {G.}~\bibnamefont {{\'E}thier-Majcher}}, \bibinfo {author}
  {\bibfnamefont {C.}~\bibnamefont {Lang}}, \bibinfo {author} {\bibfnamefont
  {E.}~\bibnamefont {Clarke}}, \bibinfo {author} {\bibfnamefont
  {M.}~\bibnamefont {Hugues}}, \bibinfo {author} {\bibfnamefont
  {C.}~\bibnamefont {Le~Gall}}, \ and\ \bibinfo {author} {\bibfnamefont
  {M.}~\bibnamefont {Atat{\"u}re}},\ }\href {\doibase
  10.1038/s41567-021-01344-7} {\bibfield  {journal} {\bibinfo  {journal} {Nat.
  Phys.}\ }\textbf {\bibinfo {volume} {17}},\ \bibinfo {pages} {1247} (\bibinfo
  {year} {2021})}\BibitemShut {NoStop}%
\bibitem [{\citenamefont {Zaporski}\ \emph {et~al.}(2023)\citenamefont
  {Zaporski}, \citenamefont {Shofer}, \citenamefont {Bodey}, \citenamefont
  {Manna}, \citenamefont {Gillard}, \citenamefont {Appel}, \citenamefont
  {Schimpf}, \citenamefont {Covre~da Silva}, \citenamefont {Jarman},
  \citenamefont {Delamare}, \citenamefont {Park}, \citenamefont {Haeusler},
  \citenamefont {Chekhovich}, \citenamefont {Rastelli}, \citenamefont
  {Gangloff}, \citenamefont {Atat{\"u}re},\ and\ \citenamefont
  {Le~Gall}}]{Zaporski:23}%
  \BibitemOpen
  \bibfield  {author} {\bibinfo {author} {\bibfnamefont {L.}~\bibnamefont
  {Zaporski}}, \bibinfo {author} {\bibfnamefont {N.}~\bibnamefont {Shofer}},
  \bibinfo {author} {\bibfnamefont {J.~H.}\ \bibnamefont {Bodey}}, \bibinfo
  {author} {\bibfnamefont {S.}~\bibnamefont {Manna}}, \bibinfo {author}
  {\bibfnamefont {G.}~\bibnamefont {Gillard}}, \bibinfo {author} {\bibfnamefont
  {M.~H.}\ \bibnamefont {Appel}}, \bibinfo {author} {\bibfnamefont
  {C.}~\bibnamefont {Schimpf}}, \bibinfo {author} {\bibfnamefont {S.~F.}\
  \bibnamefont {Covre~da Silva}}, \bibinfo {author} {\bibfnamefont
  {J.}~\bibnamefont {Jarman}}, \bibinfo {author} {\bibfnamefont
  {G.}~\bibnamefont {Delamare}}, \bibinfo {author} {\bibfnamefont
  {G.}~\bibnamefont {Park}}, \bibinfo {author} {\bibfnamefont {U.}~\bibnamefont
  {Haeusler}}, \bibinfo {author} {\bibfnamefont {E.~A.}\ \bibnamefont
  {Chekhovich}}, \bibinfo {author} {\bibfnamefont {A.}~\bibnamefont
  {Rastelli}}, \bibinfo {author} {\bibfnamefont {D.~A.}\ \bibnamefont
  {Gangloff}}, \bibinfo {author} {\bibfnamefont {M.}~\bibnamefont
  {Atat{\"u}re}}, \ and\ \bibinfo {author} {\bibfnamefont {C.}~\bibnamefont
  {Le~Gall}},\ }\href {\doibase 10.1038/s41565-022-01282-2} {\bibfield
  {journal} {\bibinfo  {journal} {Nature Nanotechnology}\ }\textbf {\bibinfo
  {volume} {18}},\ \bibinfo {pages} {257} (\bibinfo {year} {2023})}\BibitemShut
  {NoStop}%
\bibitem [{\citenamefont {Dietrich}\ \emph {et~al.}(2016)\citenamefont
  {Dietrich}, \citenamefont {Fiore}, \citenamefont {Thompson}, \citenamefont
  {Kamp},\ and\ \citenamefont {Höfling}}]{Dietrich:16}%
  \BibitemOpen
  \bibfield  {author} {\bibinfo {author} {\bibfnamefont {C.~P.}\ \bibnamefont
  {Dietrich}}, \bibinfo {author} {\bibfnamefont {A.}~\bibnamefont {Fiore}},
  \bibinfo {author} {\bibfnamefont {M.~G.}\ \bibnamefont {Thompson}}, \bibinfo
  {author} {\bibfnamefont {M.}~\bibnamefont {Kamp}}, \ and\ \bibinfo {author}
  {\bibfnamefont {S.}~\bibnamefont {Höfling}},\ }\href {\doibase
  10.1002/lpor.201500321} {\bibfield  {journal} {\bibinfo  {journal} {Laser
  Photon. Rev.}\ }\textbf {\bibinfo {volume} {10}},\ \bibinfo {pages} {870}
  (\bibinfo {year} {2016})}\BibitemShut {NoStop}%
\bibitem [{\citenamefont {Lodahl}(2018)}]{Lodahl:18}%
  \BibitemOpen
  \bibfield  {author} {\bibinfo {author} {\bibfnamefont {P.}~\bibnamefont
  {Lodahl}},\ }\href {http://stacks.iop.org/2058-9565/3/i=1/a=013001}
  {\bibfield  {journal} {\bibinfo  {journal} {Quantum Sci. Technol.}\ }\textbf
  {\bibinfo {volume} {3}},\ \bibinfo {pages} {013001} (\bibinfo {year}
  {2018})}\BibitemShut {NoStop}%
\bibitem [{\citenamefont {Grim}\ \emph {et~al.}(2019)\citenamefont {Grim},
  \citenamefont {Bracker}, \citenamefont {Zalalutdinov}, \citenamefont
  {Carter}, \citenamefont {Kozen}, \citenamefont {Kim}, \citenamefont {Kim},
  \citenamefont {Mlack}, \citenamefont {Yakes}, \citenamefont {Lee} \emph
  {et~al.}}]{Grim:19}%
  \BibitemOpen
  \bibfield  {author} {\bibinfo {author} {\bibfnamefont {J.~Q.}\ \bibnamefont
  {Grim}}, \bibinfo {author} {\bibfnamefont {A.~S.}\ \bibnamefont {Bracker}},
  \bibinfo {author} {\bibfnamefont {M.}~\bibnamefont {Zalalutdinov}}, \bibinfo
  {author} {\bibfnamefont {S.~G.}\ \bibnamefont {Carter}}, \bibinfo {author}
  {\bibfnamefont {A.~C.}\ \bibnamefont {Kozen}}, \bibinfo {author}
  {\bibfnamefont {M.}~\bibnamefont {Kim}}, \bibinfo {author} {\bibfnamefont
  {C.~S.}\ \bibnamefont {Kim}}, \bibinfo {author} {\bibfnamefont {J.~T.}\
  \bibnamefont {Mlack}}, \bibinfo {author} {\bibfnamefont {M.}~\bibnamefont
  {Yakes}}, \bibinfo {author} {\bibfnamefont {B.}~\bibnamefont {Lee}},  \emph
  {et~al.},\ }\href@noop {} {\bibfield  {journal} {\bibinfo  {journal} {Nat.
  Mater.}\ }\textbf {\bibinfo {volume} {18}},\ \bibinfo {pages} {963} (\bibinfo
  {year} {2019})}\BibitemShut {NoStop}%
\bibitem [{\citenamefont {Papon}\ \emph {et~al.}(2022)\citenamefont {Papon},
  \citenamefont {Wang}, \citenamefont {Uppu}, \citenamefont {Scholz},
  \citenamefont {Wieck}, \citenamefont {Ludwig}, \citenamefont {Lodahl},\ and\
  \citenamefont {Midolo}}]{Papon:22}%
  \BibitemOpen
  \bibfield  {author} {\bibinfo {author} {\bibfnamefont {C.}~\bibnamefont
  {Papon}}, \bibinfo {author} {\bibfnamefont {Y.}~\bibnamefont {Wang}},
  \bibinfo {author} {\bibfnamefont {R.}~\bibnamefont {Uppu}}, \bibinfo {author}
  {\bibfnamefont {S.}~\bibnamefont {Scholz}}, \bibinfo {author} {\bibfnamefont
  {A.~D.}\ \bibnamefont {Wieck}}, \bibinfo {author} {\bibfnamefont
  {A.}~\bibnamefont {Ludwig}}, \bibinfo {author} {\bibfnamefont
  {P.}~\bibnamefont {Lodahl}}, \ and\ \bibinfo {author} {\bibfnamefont
  {L.}~\bibnamefont {Midolo}},\ }\href {\doibase 10.48550/arXiv.2210.09826} {\
  ,\ \bibinfo {pages} {Preprint at https://arxiv.org/abs/2210.09826} (\bibinfo
  {year} {2022})}\BibitemShut {NoStop}%
\bibitem [{\citenamefont {Tiranov}\ \emph {et~al.}(2023)\citenamefont
  {Tiranov}, \citenamefont {Angelopoulou}, \citenamefont {van Diepen},
  \citenamefont {Schrinski}, \citenamefont {Sandberg}, \citenamefont {Wang},
  \citenamefont {Midolo}, \citenamefont {Scholz}, \citenamefont {Wieck},
  \citenamefont {Ludwig} \emph {et~al.}}]{Tiranov:23}%
  \BibitemOpen
  \bibfield  {author} {\bibinfo {author} {\bibfnamefont {A.}~\bibnamefont
  {Tiranov}}, \bibinfo {author} {\bibfnamefont {V.}~\bibnamefont
  {Angelopoulou}}, \bibinfo {author} {\bibfnamefont {C.~J.}\ \bibnamefont {van
  Diepen}}, \bibinfo {author} {\bibfnamefont {B.}~\bibnamefont {Schrinski}},
  \bibinfo {author} {\bibfnamefont {O.~A.~D.}\ \bibnamefont {Sandberg}},
  \bibinfo {author} {\bibfnamefont {Y.}~\bibnamefont {Wang}}, \bibinfo {author}
  {\bibfnamefont {L.}~\bibnamefont {Midolo}}, \bibinfo {author} {\bibfnamefont
  {S.}~\bibnamefont {Scholz}}, \bibinfo {author} {\bibfnamefont {A.~D.}\
  \bibnamefont {Wieck}}, \bibinfo {author} {\bibfnamefont {A.}~\bibnamefont
  {Ludwig}},  \emph {et~al.},\ }\href {\doibase 10.1126/science.ade9324}
  {\bibfield  {journal} {\bibinfo  {journal} {Science}\ }\textbf {\bibinfo
  {volume} {379}},\ \bibinfo {pages} {389} (\bibinfo {year}
  {2023})}\BibitemShut {NoStop}%
\bibitem [{\citenamefont {J{\"o}ns}\ \emph {et~al.}(2015)\citenamefont
  {J{\"o}ns}, \citenamefont {Rengstl}, \citenamefont {Oster}, \citenamefont
  {Hargart}, \citenamefont {Heldmaier}, \citenamefont {Bounouar}, \citenamefont
  {Ulrich}, \citenamefont {Jetter},\ and\ \citenamefont {Michler}}]{Jons:15}%
  \BibitemOpen
  \bibfield  {author} {\bibinfo {author} {\bibfnamefont {K.~D.}\ \bibnamefont
  {J{\"o}ns}}, \bibinfo {author} {\bibfnamefont {U.}~\bibnamefont {Rengstl}},
  \bibinfo {author} {\bibfnamefont {M.}~\bibnamefont {Oster}}, \bibinfo
  {author} {\bibfnamefont {F.}~\bibnamefont {Hargart}}, \bibinfo {author}
  {\bibfnamefont {M.}~\bibnamefont {Heldmaier}}, \bibinfo {author}
  {\bibfnamefont {S.}~\bibnamefont {Bounouar}}, \bibinfo {author}
  {\bibfnamefont {S.~M.}\ \bibnamefont {Ulrich}}, \bibinfo {author}
  {\bibfnamefont {M.}~\bibnamefont {Jetter}}, \ and\ \bibinfo {author}
  {\bibfnamefont {P.}~\bibnamefont {Michler}},\ }\href {\doibase
  10.1088/0022-3727/48/8/085101} {\bibfield  {journal} {\bibinfo  {journal} {J.
  Phys. D: Appl. Phys}\ }\textbf {\bibinfo {volume} {48}},\ \bibinfo {pages}
  {085101} (\bibinfo {year} {2015})}\BibitemShut {NoStop}%
\bibitem [{\citenamefont {Kim}\ \emph {et~al.}(2020)\citenamefont {Kim},
  \citenamefont {Aghaeimeibodi}, \citenamefont {Carolan}, \citenamefont
  {Englund},\ and\ \citenamefont {Waks}}]{Kim:20}%
  \BibitemOpen
  \bibfield  {author} {\bibinfo {author} {\bibfnamefont {J.-H.}\ \bibnamefont
  {Kim}}, \bibinfo {author} {\bibfnamefont {S.}~\bibnamefont {Aghaeimeibodi}},
  \bibinfo {author} {\bibfnamefont {J.}~\bibnamefont {Carolan}}, \bibinfo
  {author} {\bibfnamefont {D.}~\bibnamefont {Englund}}, \ and\ \bibinfo
  {author} {\bibfnamefont {E.}~\bibnamefont {Waks}},\ }\href {\doibase
  10.1364/OPTICA.384118} {\bibfield  {journal} {\bibinfo  {journal} {Optica}\
  }\textbf {\bibinfo {volume} {7}},\ \bibinfo {pages} {291} (\bibinfo {year}
  {2020})}\BibitemShut {NoStop}%
\bibitem [{\citenamefont {Elshaari}\ \emph {et~al.}(2020)\citenamefont
  {Elshaari}, \citenamefont {Pernice}, \citenamefont {Srinivasan},
  \citenamefont {Benson},\ and\ \citenamefont {Zwiller}}]{Elshaari:20}%
  \BibitemOpen
  \bibfield  {author} {\bibinfo {author} {\bibfnamefont {A.~W.}\ \bibnamefont
  {Elshaari}}, \bibinfo {author} {\bibfnamefont {W.}~\bibnamefont {Pernice}},
  \bibinfo {author} {\bibfnamefont {K.}~\bibnamefont {Srinivasan}}, \bibinfo
  {author} {\bibfnamefont {O.}~\bibnamefont {Benson}}, \ and\ \bibinfo {author}
  {\bibfnamefont {V.}~\bibnamefont {Zwiller}},\ }\href {\doibase
  10.1038/s41566-020-0609-x} {\bibfield  {journal} {\bibinfo  {journal} {Nat.
  Photon.}\ }\textbf {\bibinfo {volume} {14}},\ \bibinfo {pages} {285}
  (\bibinfo {year} {2020})}\BibitemShut {NoStop}%
\bibitem [{\citenamefont {Kim}\ \emph {et~al.}(2017)\citenamefont {Kim},
  \citenamefont {Aghaeimeibodi}, \citenamefont {Richardson}, \citenamefont
  {Leavitt}, \citenamefont {Englund},\ and\ \citenamefont {Waks}}]{Kim:17}%
  \BibitemOpen
  \bibfield  {author} {\bibinfo {author} {\bibfnamefont {J.-H.}\ \bibnamefont
  {Kim}}, \bibinfo {author} {\bibfnamefont {S.}~\bibnamefont {Aghaeimeibodi}},
  \bibinfo {author} {\bibfnamefont {C.~J.}\ \bibnamefont {Richardson}},
  \bibinfo {author} {\bibfnamefont {R.~P.}\ \bibnamefont {Leavitt}}, \bibinfo
  {author} {\bibfnamefont {D.}~\bibnamefont {Englund}}, \ and\ \bibinfo
  {author} {\bibfnamefont {E.}~\bibnamefont {Waks}},\ }\href {\doibase
  10.1021/acs.nanolett.7b03220} {\bibfield  {journal} {\bibinfo  {journal}
  {Nano Lett.}\ }\textbf {\bibinfo {volume} {17}},\ \bibinfo {pages} {7394}
  (\bibinfo {year} {2017})}\BibitemShut {NoStop}%
\bibitem [{\citenamefont {Katsumi}\ \emph {et~al.}(2019)\citenamefont
  {Katsumi}, \citenamefont {Ota}, \citenamefont {Osada}, \citenamefont
  {Yamaguchi}, \citenamefont {Tajiri}, \citenamefont {Kakuda}, \citenamefont
  {Iwamoto}, \citenamefont {Akiyama},\ and\ \citenamefont
  {Arakawa}}]{Katsumi:19}%
  \BibitemOpen
  \bibfield  {author} {\bibinfo {author} {\bibfnamefont {R.}~\bibnamefont
  {Katsumi}}, \bibinfo {author} {\bibfnamefont {Y.}~\bibnamefont {Ota}},
  \bibinfo {author} {\bibfnamefont {A.}~\bibnamefont {Osada}}, \bibinfo
  {author} {\bibfnamefont {T.}~\bibnamefont {Yamaguchi}}, \bibinfo {author}
  {\bibfnamefont {T.}~\bibnamefont {Tajiri}}, \bibinfo {author} {\bibfnamefont
  {M.}~\bibnamefont {Kakuda}}, \bibinfo {author} {\bibfnamefont
  {S.}~\bibnamefont {Iwamoto}}, \bibinfo {author} {\bibfnamefont
  {H.}~\bibnamefont {Akiyama}}, \ and\ \bibinfo {author} {\bibfnamefont
  {Y.}~\bibnamefont {Arakawa}},\ }\href {\doibase 10.1063/1.5087263} {\bibfield
   {journal} {\bibinfo  {journal} {APL Photonics}\ }\textbf {\bibinfo {volume}
  {4}},\ \bibinfo {pages} {036105} (\bibinfo {year} {2019})}\BibitemShut
  {NoStop}%
\bibitem [{\citenamefont {Katsumi}\ \emph {et~al.}(2020)\citenamefont
  {Katsumi}, \citenamefont {Ota}, \citenamefont {Osada}, \citenamefont
  {Tajiri}, \citenamefont {Yamaguchi}, \citenamefont {Kakuda}, \citenamefont
  {Iwamoto}, \citenamefont {Akiyama},\ and\ \citenamefont
  {Arakawa}}]{Katsumi:20}%
  \BibitemOpen
  \bibfield  {author} {\bibinfo {author} {\bibfnamefont {R.}~\bibnamefont
  {Katsumi}}, \bibinfo {author} {\bibfnamefont {Y.}~\bibnamefont {Ota}},
  \bibinfo {author} {\bibfnamefont {A.}~\bibnamefont {Osada}}, \bibinfo
  {author} {\bibfnamefont {T.}~\bibnamefont {Tajiri}}, \bibinfo {author}
  {\bibfnamefont {T.}~\bibnamefont {Yamaguchi}}, \bibinfo {author}
  {\bibfnamefont {M.}~\bibnamefont {Kakuda}}, \bibinfo {author} {\bibfnamefont
  {S.}~\bibnamefont {Iwamoto}}, \bibinfo {author} {\bibfnamefont
  {H.}~\bibnamefont {Akiyama}}, \ and\ \bibinfo {author} {\bibfnamefont
  {Y.}~\bibnamefont {Arakawa}},\ }\href {\doibase 10.1063/1.5129325} {\bibfield
   {journal} {\bibinfo  {journal} {Appl. Phys. Lett.}\ }\textbf {\bibinfo
  {volume} {116}},\ \bibinfo {pages} {041103} (\bibinfo {year}
  {2020})}\BibitemShut {NoStop}%
\bibitem [{\citenamefont {Mouradian}\ \emph {et~al.}(2015)\citenamefont
  {Mouradian}, \citenamefont {Schr\"oder}, \citenamefont {Poitras},
  \citenamefont {Li}, \citenamefont {Goldstein}, \citenamefont {Chen},
  \citenamefont {Walsh}, \citenamefont {Cardenas}, \citenamefont {Markham},
  \citenamefont {Twitchen}, \citenamefont {Lipson},\ and\ \citenamefont
  {Englund}}]{Mouradian:15}%
  \BibitemOpen
  \bibfield  {author} {\bibinfo {author} {\bibfnamefont {S.~L.}\ \bibnamefont
  {Mouradian}}, \bibinfo {author} {\bibfnamefont {T.}~\bibnamefont
  {Schr\"oder}}, \bibinfo {author} {\bibfnamefont {C.~B.}\ \bibnamefont
  {Poitras}}, \bibinfo {author} {\bibfnamefont {L.}~\bibnamefont {Li}},
  \bibinfo {author} {\bibfnamefont {J.}~\bibnamefont {Goldstein}}, \bibinfo
  {author} {\bibfnamefont {E.~H.}\ \bibnamefont {Chen}}, \bibinfo {author}
  {\bibfnamefont {M.}~\bibnamefont {Walsh}}, \bibinfo {author} {\bibfnamefont
  {J.}~\bibnamefont {Cardenas}}, \bibinfo {author} {\bibfnamefont {M.~L.}\
  \bibnamefont {Markham}}, \bibinfo {author} {\bibfnamefont {D.~J.}\
  \bibnamefont {Twitchen}}, \bibinfo {author} {\bibfnamefont {M.}~\bibnamefont
  {Lipson}}, \ and\ \bibinfo {author} {\bibfnamefont {D.}~\bibnamefont
  {Englund}},\ }\href {\doibase 10.1103/PhysRevX.5.031009} {\bibfield
  {journal} {\bibinfo  {journal} {Phys. Rev. X}\ }\textbf {\bibinfo {volume}
  {5}},\ \bibinfo {pages} {031009} (\bibinfo {year} {2015})}\BibitemShut
  {NoStop}%
\bibitem [{\citenamefont {Zadeh}\ \emph {et~al.}(2016)\citenamefont {Zadeh},
  \citenamefont {Elshaari}, \citenamefont {J{\"o}ns}, \citenamefont {Fognini},
  \citenamefont {Dalacu}, \citenamefont {Poole}, \citenamefont {Reimer},\ and\
  \citenamefont {Zwiller}}]{Zadeh:16}%
  \BibitemOpen
  \bibfield  {author} {\bibinfo {author} {\bibfnamefont {I.~E.}\ \bibnamefont
  {Zadeh}}, \bibinfo {author} {\bibfnamefont {A.~W.}\ \bibnamefont {Elshaari}},
  \bibinfo {author} {\bibfnamefont {K.~D.}\ \bibnamefont {J{\"o}ns}}, \bibinfo
  {author} {\bibfnamefont {A.}~\bibnamefont {Fognini}}, \bibinfo {author}
  {\bibfnamefont {D.}~\bibnamefont {Dalacu}}, \bibinfo {author} {\bibfnamefont
  {P.~J.}\ \bibnamefont {Poole}}, \bibinfo {author} {\bibfnamefont {M.~E.}\
  \bibnamefont {Reimer}}, \ and\ \bibinfo {author} {\bibfnamefont
  {V.}~\bibnamefont {Zwiller}},\ }\href {\doibase 10.1021/acs.nanolett.5b04709}
  {\bibfield  {journal} {\bibinfo  {journal} {Nano Lett.}\ }\textbf {\bibinfo
  {volume} {16}},\ \bibinfo {pages} {2289} (\bibinfo {year}
  {2016})}
\BibitemShut {NoStop}%
\bibitem [{\citenamefont {Elshaari}\ \emph {et~al.}(2017)\citenamefont
  {Elshaari}, \citenamefont {Zadeh}, \citenamefont {Fognini}, \citenamefont
  {Reimer}, \citenamefont {Dalacu}, \citenamefont {Poole}, \citenamefont
  {Zwiller},\ and\ \citenamefont {J{\"o}ns}}]{Elshaari:17}%
  \BibitemOpen
  \bibfield  {author} {\bibinfo {author} {\bibfnamefont {A.~W.}\ \bibnamefont
  {Elshaari}}, \bibinfo {author} {\bibfnamefont {I.~E.}\ \bibnamefont {Zadeh}},
  \bibinfo {author} {\bibfnamefont {A.}~\bibnamefont {Fognini}}, \bibinfo
  {author} {\bibfnamefont {M.~E.}\ \bibnamefont {Reimer}}, \bibinfo {author}
  {\bibfnamefont {D.}~\bibnamefont {Dalacu}}, \bibinfo {author} {\bibfnamefont
  {P.~J.}\ \bibnamefont {Poole}}, \bibinfo {author} {\bibfnamefont
  {V.}~\bibnamefont {Zwiller}}, \ and\ \bibinfo {author} {\bibfnamefont
  {K.~D.}\ \bibnamefont {J{\"o}ns}},\ }\href {\doibase
  10.1038/s41467-017-00486-8} {\bibfield  {journal} {\bibinfo  {journal} {Nat.
  Commun.}\ }\textbf {\bibinfo {volume} {8}},\ \bibinfo {pages} {379} (\bibinfo
  {year} {2017})} \BibitemShut {NoStop}%
\bibitem [{\citenamefont {Davanco}\ \emph {et~al.}(2017)\citenamefont
  {Davanco}, \citenamefont {Liu}, \citenamefont {Sapienza}, \citenamefont
  {Zhang}, \citenamefont {De~Miranda~Cardoso}, \citenamefont {Verma},
  \citenamefont {Mirin}, \citenamefont {Nam}, \citenamefont {Liu},\ and\
  \citenamefont {Srinivasan}}]{Davanco:17}%
  \BibitemOpen
  \bibfield  {author} {\bibinfo {author} {\bibfnamefont {M.}~\bibnamefont
  {Davanco}}, \bibinfo {author} {\bibfnamefont {J.}~\bibnamefont {Liu}},
  \bibinfo {author} {\bibfnamefont {L.}~\bibnamefont {Sapienza}}, \bibinfo
  {author} {\bibfnamefont {C.-Z.}\ \bibnamefont {Zhang}}, \bibinfo {author}
  {\bibfnamefont {J.}~\bibnamefont {De~Miranda~Cardoso}}, \bibinfo {author}
  {\bibfnamefont {V.}~\bibnamefont {Verma}}, \bibinfo {author} {\bibfnamefont
  {R.}~\bibnamefont {Mirin}}, \bibinfo {author} {\bibfnamefont {S.~W.}\
  \bibnamefont {Nam}}, \bibinfo {author} {\bibfnamefont {L.}~\bibnamefont
  {Liu}}, \ and\ \bibinfo {author} {\bibfnamefont {K.}~\bibnamefont
  {Srinivasan}},\ }\href {\doibase 10.1038/s41467-017-00987-6} {\bibfield
  {journal} {\bibinfo  {journal} {Nat. Commun.}\ }\textbf {\bibinfo {volume}
  {8}},\ \bibinfo {pages} {889} (\bibinfo {year} {2017})}\BibitemShut {NoStop}%
\bibitem [{\citenamefont {Elshaari}\ \emph {et~al.}(2018)\citenamefont
  {Elshaari}, \citenamefont {B{\"u}y{\"u}k{\"o}zer}, \citenamefont {Zadeh},
  \citenamefont {Lettner}, \citenamefont {Zhao}, \citenamefont {Sch{\"o}ll},
  \citenamefont {Gyger}, \citenamefont {Reimer}, \citenamefont {Dalacu},
  \citenamefont {Poole} \emph {et~al.}}]{Elshaari:18}%
  \BibitemOpen
  \bibfield  {author} {\bibinfo {author} {\bibfnamefont {A.~W.}\ \bibnamefont
  {Elshaari}}, \bibinfo {author} {\bibfnamefont {E.}~\bibnamefont
  {B{\"u}y{\"u}k{\"o}zer}}, \bibinfo {author} {\bibfnamefont {I.~E.}\ \bibnamefont
  {Zadeh}}, \bibinfo {author} {\bibfnamefont {T.}~\bibnamefont {Lettner}},
  \bibinfo {author} {\bibfnamefont {P.}~\bibnamefont {Zhao}}, \bibinfo {author}
  {\bibfnamefont {E.}~\bibnamefont {Sch{\"o}ll}}, \bibinfo {author}
  {\bibfnamefont {S.}~\bibnamefont {Gyger}}, \bibinfo {author} {\bibfnamefont
  {M.~E.}\ \bibnamefont {Reimer}}, \bibinfo {author} {\bibfnamefont
  {D.}~\bibnamefont {Dalacu}}, \bibinfo {author} {\bibfnamefont {P.~J.}\
  \bibnamefont {Poole}},  \emph {et~al.},\ }\href {\doibase
  10.1021/acs.nanolett.8b03937} {\bibfield  {journal} {\bibinfo  {journal}
  {Nano Lett.}\ }\textbf {\bibinfo {volume} {18}},\ \bibinfo {pages} {7969}
  (\bibinfo {year} {2018})}\BibitemShut {NoStop}%
\bibitem [{\citenamefont {Chanana}\ \emph {et~al.}(2022)\citenamefont
  {Chanana}, \citenamefont {Larocque}, \citenamefont {Moreira}, \citenamefont
  {Carolan}, \citenamefont {Guha}, \citenamefont {Melo}, \citenamefont {Anant},
  \citenamefont {Song}, \citenamefont {Englund}, \citenamefont {Blumenthal},
  \citenamefont {Srinivasan},\ and\ \citenamefont {Davanco}}]{Chanana:22}%
  \BibitemOpen
  \bibfield  {author} {\bibinfo {author} {\bibfnamefont {A.}~\bibnamefont
  {Chanana}}, \bibinfo {author} {\bibfnamefont {H.}~\bibnamefont {Larocque}},
  \bibinfo {author} {\bibfnamefont {R.}~\bibnamefont {Moreira}}, \bibinfo
  {author} {\bibfnamefont {J.}~\bibnamefont {Carolan}}, \bibinfo {author}
  {\bibfnamefont {B.}~\bibnamefont {Guha}}, \bibinfo {author} {\bibfnamefont
  {E.~G.}\ \bibnamefont {Melo}}, \bibinfo {author} {\bibfnamefont
  {V.}~\bibnamefont {Anant}}, \bibinfo {author} {\bibfnamefont
  {J.}~\bibnamefont {Song}}, \bibinfo {author} {\bibfnamefont {D.}~\bibnamefont
  {Englund}}, \bibinfo {author} {\bibfnamefont {D.~J.}\ \bibnamefont
  {Blumenthal}}, \bibinfo {author} {\bibfnamefont {K.}~\bibnamefont
  {Srinivasan}}, \ and\ \bibinfo {author} {\bibfnamefont {M.}~\bibnamefont
  {Davanco}},\ }\href {\doibase 10.1038/s41467-022-35332-z} {\bibfield
  {journal} {\bibinfo  {journal} {Nat. Commun.}\ }\textbf {\bibinfo {volume}
  {13}},\ \bibinfo {pages} {7693} (\bibinfo {year} {2022})}\BibitemShut
  {NoStop}%
\bibitem [{\citenamefont {Sun}\ \emph {et~al.}(2013)\citenamefont {Sun},
  \citenamefont {Timurdogan}, \citenamefont {Yaacobi}, \citenamefont
  {Hosseini},\ and\ \citenamefont {Watts}}]{Sun:13}%
  \BibitemOpen
  \bibfield  {author} {\bibinfo {author} {\bibfnamefont {J.}~\bibnamefont
  {Sun}}, \bibinfo {author} {\bibfnamefont {E.}~\bibnamefont {Timurdogan}},
  \bibinfo {author} {\bibfnamefont {A.}~\bibnamefont {Yaacobi}}, \bibinfo
  {author} {\bibfnamefont {E.~S.}\ \bibnamefont {Hosseini}}, \ and\ \bibinfo
  {author} {\bibfnamefont {M.~R.}\ \bibnamefont {Watts}},\ }\href {\doibase
  10.1038/nature11727} {\bibfield  {journal} {\bibinfo  {journal} {Nature}\
  }\textbf {\bibinfo {volume} {493}},\ \bibinfo {pages} {195} (\bibinfo {year}
  {2013})}\BibitemShut {NoStop}%
\bibitem [{\citenamefont {Bogaerts}\ \emph
  {et~al.}(2020{\natexlab{a}})\citenamefont {Bogaerts}, \citenamefont
  {P{\'e}rez}, \citenamefont {Capmany}, \citenamefont {Miller}, \citenamefont
  {Poon}, \citenamefont {Englund}, \citenamefont {Morichetti},\ and\
  \citenamefont {Melloni}}]{Bogaerts:20}%
  \BibitemOpen
  \bibfield  {author} {\bibinfo {author} {\bibfnamefont {W.}~\bibnamefont
  {Bogaerts}}, \bibinfo {author} {\bibfnamefont {D.}~\bibnamefont {P{\'e}rez}},
  \bibinfo {author} {\bibfnamefont {J.}~\bibnamefont {Capmany}}, \bibinfo
  {author} {\bibfnamefont {D.~A.}\ \bibnamefont {Miller}}, \bibinfo {author}
  {\bibfnamefont {J.}~\bibnamefont {Poon}}, \bibinfo {author} {\bibfnamefont
  {D.}~\bibnamefont {Englund}}, \bibinfo {author} {\bibfnamefont
  {F.}~\bibnamefont {Morichetti}}, \ and\ \bibinfo {author} {\bibfnamefont
  {A.}~\bibnamefont {Melloni}},\ }\href {\doibase 10.1038/s41586-020-2764-0}
  {\bibfield  {journal} {\bibinfo  {journal} {Nature}\ }\textbf {\bibinfo
  {volume} {586}},\ \bibinfo {pages} {207} (\bibinfo {year}
  {2020}{\natexlab{a}})}\BibitemShut {NoStop}%
\bibitem [{\citenamefont {Bogaerts}\ \emph
  {et~al.}(2020{\natexlab{b}})\citenamefont {Bogaerts}, \citenamefont
  {Sattari}, \citenamefont {Edinger}, \citenamefont {Takabayashi},
  \citenamefont {Zand}, \citenamefont {Wang}, \citenamefont {Ribeiro},
  \citenamefont {Jezzini}, \citenamefont {Errando-Herranz}, \citenamefont
  {Talli} \emph {et~al.}}]{Bogaerts:20b}%
  \BibitemOpen
  \bibfield  {author} {\bibinfo {author} {\bibfnamefont {W.}~\bibnamefont
  {Bogaerts}}, \bibinfo {author} {\bibfnamefont {H.}~\bibnamefont {Sattari}},
  \bibinfo {author} {\bibfnamefont {P.}~\bibnamefont {Edinger}}, \bibinfo
  {author} {\bibfnamefont {A.~Y.}\ \bibnamefont {Takabayashi}}, \bibinfo
  {author} {\bibfnamefont {I.}~\bibnamefont {Zand}}, \bibinfo {author}
  {\bibfnamefont {X.}~\bibnamefont {Wang}}, \bibinfo {author} {\bibfnamefont
  {A.}~\bibnamefont {Ribeiro}}, \bibinfo {author} {\bibfnamefont
  {M.}~\bibnamefont {Jezzini}}, \bibinfo {author} {\bibfnamefont
  {C.}~\bibnamefont {Errando-Herranz}}, \bibinfo {author} {\bibfnamefont
  {G.}~\bibnamefont {Talli}},  \emph {et~al.},\ }in\ \href {\doibase
  10.1117/12.2540934} {\emph {\bibinfo {booktitle} {Silicon Photonics XV}}},\
  Vol.\ \bibinfo {volume} {11285}\ (\bibinfo {organization} {SPIE},\ \bibinfo
  {year} {2020})\ p.\ \bibinfo {pages} {1128503}\BibitemShut {NoStop}%
\bibitem [{\citenamefont {Edinger}\ \emph {et~al.}(2021)\citenamefont
  {Edinger}, \citenamefont {Takabayashi}, \citenamefont {Errando-Herranz},
  \citenamefont {Khan}, \citenamefont {Sattari}, \citenamefont {Verheyen},
  \citenamefont {Bogaerts}, \citenamefont {Quack},\ and\ \citenamefont
  {Gylfason}}]{Edinger:21}%
  \BibitemOpen
  \bibfield  {author} {\bibinfo {author} {\bibfnamefont {P.}~\bibnamefont
  {Edinger}}, \bibinfo {author} {\bibfnamefont {A.~Y.}\ \bibnamefont
  {Takabayashi}}, \bibinfo {author} {\bibfnamefont {C.}~\bibnamefont
  {Errando-Herranz}}, \bibinfo {author} {\bibfnamefont {U.}~\bibnamefont
  {Khan}}, \bibinfo {author} {\bibfnamefont {H.}~\bibnamefont {Sattari}},
  \bibinfo {author} {\bibfnamefont {P.}~\bibnamefont {Verheyen}}, \bibinfo
  {author} {\bibfnamefont {W.}~\bibnamefont {Bogaerts}}, \bibinfo {author}
  {\bibfnamefont {N.}~\bibnamefont {Quack}}, \ and\ \bibinfo {author}
  {\bibfnamefont {K.~B.}\ \bibnamefont {Gylfason}},\ }\href {\doibase
  10.1364/OL.436288} {\bibfield  {journal} {\bibinfo  {journal} {Opt. Lett.}\
  }\textbf {\bibinfo {volume} {46}},\ \bibinfo {pages} {5671} (\bibinfo {year}
  {2021})}\BibitemShut {NoStop}%
\bibitem [{\citenamefont {Miller}\ \emph {et~al.}(1984)\citenamefont {Miller},
  \citenamefont {Chemla}, \citenamefont {Damen}, \citenamefont {Gossard},
  \citenamefont {Wiegmann}, \citenamefont {Wood},\ and\ \citenamefont
  {Burrus}}]{Miller:84i}%
  \BibitemOpen
  \bibfield  {author} {\bibinfo {author} {\bibfnamefont {D.~A.~B.}\
  \bibnamefont {Miller}}, \bibinfo {author} {\bibfnamefont {D.~S.}\
  \bibnamefont {Chemla}}, \bibinfo {author} {\bibfnamefont {T.~C.}\
  \bibnamefont {Damen}}, \bibinfo {author} {\bibfnamefont {A.~C.}\ \bibnamefont
  {Gossard}}, \bibinfo {author} {\bibfnamefont {W.}~\bibnamefont {Wiegmann}},
  \bibinfo {author} {\bibfnamefont {T.~H.}\ \bibnamefont {Wood}}, \ and\
  \bibinfo {author} {\bibfnamefont {C.~A.}\ \bibnamefont {Burrus}},\ }\href
  {\doibase 10.1103/PhysRevLett.53.2173} {\bibfield  {journal} {\bibinfo
  {journal} {Phys. Rev. Lett.}\ }\textbf {\bibinfo {volume} {53}},\ \bibinfo
  {pages} {2173} (\bibinfo {year} {1984})}\BibitemShut {NoStop}%
\bibitem [{\citenamefont {Miller}\ \emph {et~al.}(1985)\citenamefont {Miller},
  \citenamefont {Chemla}, \citenamefont {Damen}, \citenamefont {Gossard},
  \citenamefont {Wiegmann}, \citenamefont {Wood},\ and\ \citenamefont
  {Burrus}}]{Miller:84ii}%
  \BibitemOpen
  \bibfield  {author} {\bibinfo {author} {\bibfnamefont {D.~A.~B.}\
  \bibnamefont {Miller}}, \bibinfo {author} {\bibfnamefont {D.~S.}\
  \bibnamefont {Chemla}}, \bibinfo {author} {\bibfnamefont {T.~C.}\
  \bibnamefont {Damen}}, \bibinfo {author} {\bibfnamefont {A.~C.}\ \bibnamefont
  {Gossard}}, \bibinfo {author} {\bibfnamefont {W.}~\bibnamefont {Wiegmann}},
  \bibinfo {author} {\bibfnamefont {T.~H.}\ \bibnamefont {Wood}}, \ and\
  \bibinfo {author} {\bibfnamefont {C.~A.}\ \bibnamefont {Burrus}},\ }\href
  {\doibase 10.1103/PhysRevB.32.1043} {\bibfield  {journal} {\bibinfo
  {journal} {Phys. Rev. B}\ }\textbf {\bibinfo {volume} {32}},\ \bibinfo
  {pages} {1043} (\bibinfo {year} {1985})}\BibitemShut {NoStop}%
\bibitem [{\citenamefont {Timurdogan}\ \emph {et~al.}(2019)\citenamefont
  {Timurdogan}, \citenamefont {Su}, \citenamefont {Shiue}, \citenamefont
  {Poulton}, \citenamefont {Byrd}, \citenamefont {Xin},\ and\ \citenamefont
  {Watts}}]{Timurdogan:19}%
  \BibitemOpen
  \bibfield  {author} {\bibinfo {author} {\bibfnamefont {E.}~\bibnamefont
  {Timurdogan}}, \bibinfo {author} {\bibfnamefont {Z.}~\bibnamefont {Su}},
  \bibinfo {author} {\bibfnamefont {R.-J.}\ \bibnamefont {Shiue}}, \bibinfo
  {author} {\bibfnamefont {C.~V.}\ \bibnamefont {Poulton}}, \bibinfo {author}
  {\bibfnamefont {M.~J.}\ \bibnamefont {Byrd}}, \bibinfo {author}
  {\bibfnamefont {S.}~\bibnamefont {Xin}}, \ and\ \bibinfo {author}
  {\bibfnamefont {M.~R.}\ \bibnamefont {Watts}},\ }in\ \href {\doibase
  10.1364/OFC.2019.Tu2A.1} {\emph {\bibinfo {booktitle} {Optical Fiber
  Communication Conference (OFC) 2019}}}\ (\bibinfo  {publisher} {Optica
  Publishing Group},\ \bibinfo {year} {2019})\ p.\ \bibinfo {pages}
  {Tu2A.1}\BibitemShut {NoStop}%
\bibitem [{\citenamefont {Lee}\ \emph {et~al.}(2020)\citenamefont {Lee},
  \citenamefont {Buyukkaya}, \citenamefont {Harper}, \citenamefont
  {Aghaeimeibodi}, \citenamefont {Richardson},\ and\ \citenamefont
  {Waks}}]{Lee:20}%
  \BibitemOpen
  \bibfield  {author} {\bibinfo {author} {\bibfnamefont {C.-M.}\ \bibnamefont
  {Lee}}, \bibinfo {author} {\bibfnamefont {M.~A.}\ \bibnamefont {Buyukkaya}},
  \bibinfo {author} {\bibfnamefont {S.}~\bibnamefont {Harper}}, \bibinfo
  {author} {\bibfnamefont {S.}~\bibnamefont {Aghaeimeibodi}}, \bibinfo {author}
  {\bibfnamefont {C.~J.}\ \bibnamefont {Richardson}}, \ and\ \bibinfo {author}
  {\bibfnamefont {E.}~\bibnamefont {Waks}},\ }\href {\doibase
  10.1021/acs.nanolett.0c03680} {\bibfield  {journal} {\bibinfo  {journal}
  {Nano Lett.}\ }\textbf {\bibinfo {volume} {21}},\ \bibinfo {pages} {323}
  (\bibinfo {year} {2020})}\BibitemShut {NoStop}%
\bibitem [{\citenamefont {Aghaeimeibodi}\ \emph {et~al.}(2019)\citenamefont
  {Aghaeimeibodi}, \citenamefont {Lee}, \citenamefont {Buyukkaya},
  \citenamefont {Richardson},\ and\ \citenamefont {Waks}}]{Aghaeimeibodi:19}%
  \BibitemOpen
  \bibfield  {author} {\bibinfo {author} {\bibfnamefont {S.}~\bibnamefont
  {Aghaeimeibodi}}, \bibinfo {author} {\bibfnamefont {C.-M.}\ \bibnamefont
  {Lee}}, \bibinfo {author} {\bibfnamefont {M.~A.}\ \bibnamefont {Buyukkaya}},
  \bibinfo {author} {\bibfnamefont {C.~J.}\ \bibnamefont {Richardson}}, \ and\
  \bibinfo {author} {\bibfnamefont {E.}~\bibnamefont {Waks}},\ }\href {\doibase
  10.1063/1.5082560} {\bibfield  {journal} {\bibinfo  {journal} {Appl. Phys.
  Lett.}\ }\textbf {\bibinfo {volume} {114}},\ \bibinfo {pages} {071105}
  (\bibinfo {year} {2019})}\BibitemShut {NoStop}%
\bibitem [{\citenamefont {Muller}\ \emph {et~al.}(2007)\citenamefont {Muller},
  \citenamefont {Flagg}, \citenamefont {Bianucci}, \citenamefont {Wang},
  \citenamefont {Deppe}, \citenamefont {Ma}, \citenamefont {Zhang},
  \citenamefont {Salamo}, \citenamefont {Xiao},\ and\ \citenamefont
  {Shih}}]{Muller:07}%
  \BibitemOpen
  \bibfield  {author} {\bibinfo {author} {\bibfnamefont {A.}~\bibnamefont
  {Muller}}, \bibinfo {author} {\bibfnamefont {E.~B.}\ \bibnamefont {Flagg}},
  \bibinfo {author} {\bibfnamefont {P.}~\bibnamefont {Bianucci}}, \bibinfo
  {author} {\bibfnamefont {X.~Y.}\ \bibnamefont {Wang}}, \bibinfo {author}
  {\bibfnamefont {D.~G.}\ \bibnamefont {Deppe}}, \bibinfo {author}
  {\bibfnamefont {W.}~\bibnamefont {Ma}}, \bibinfo {author} {\bibfnamefont
  {J.}~\bibnamefont {Zhang}}, \bibinfo {author} {\bibfnamefont {G.~J.}\
  \bibnamefont {Salamo}}, \bibinfo {author} {\bibfnamefont {M.}~\bibnamefont
  {Xiao}}, \ and\ \bibinfo {author} {\bibfnamefont {C.~K.}\ \bibnamefont
  {Shih}},\ }\href {\doibase 10.1103/PhysRevLett.99.187402} {\bibfield
  {journal} {\bibinfo  {journal} {Phys. Rev. Lett.}\ }\textbf {\bibinfo
  {volume} {99}},\ \bibinfo {pages} {187402} (\bibinfo {year}
  {2007})}\BibitemShut {NoStop}%
\bibitem [{\citenamefont {Huber}\ \emph {et~al.}(2020)\citenamefont {Huber},
  \citenamefont {Davanco}, \citenamefont {M\"{u}ller}, \citenamefont {Shuai},
  \citenamefont {Gazzano},\ and\ \citenamefont {Solomon}}]{Huber:20}%
  \BibitemOpen
  \bibfield  {author} {\bibinfo {author} {\bibfnamefont {T.}~\bibnamefont
  {Huber}}, \bibinfo {author} {\bibfnamefont {M.}~\bibnamefont {Davanco}},
  \bibinfo {author} {\bibfnamefont {M.}~\bibnamefont {M\"{u}ller}}, \bibinfo
  {author} {\bibfnamefont {Y.}~\bibnamefont {Shuai}}, \bibinfo {author}
  {\bibfnamefont {O.}~\bibnamefont {Gazzano}}, \ and\ \bibinfo {author}
  {\bibfnamefont {G.~S.}\ \bibnamefont {Solomon}},\ }\href {\doibase
  10.1364/OPTICA.382273} {\bibfield  {journal} {\bibinfo  {journal} {Optica}\
  }\textbf {\bibinfo {volume} {7}},\ \bibinfo {pages} {380} (\bibinfo {year}
  {2020})}\BibitemShut {NoStop}%
\bibitem [{\citenamefont {Reithmaier}\ \emph {et~al.}(2015)\citenamefont
  {Reithmaier}, \citenamefont {Kaniber}, \citenamefont {Flassig}, \citenamefont
  {Lichtmannecker}, \citenamefont {M{\"u}ller}, \citenamefont {Andrejew},
  \citenamefont {Vuckovic}, \citenamefont {Gross},\ and\ \citenamefont
  {Finley}}]{Reithmaier:15}%
  \BibitemOpen
  \bibfield  {author} {\bibinfo {author} {\bibfnamefont {G.}~\bibnamefont
  {Reithmaier}}, \bibinfo {author} {\bibfnamefont {M.}~\bibnamefont {Kaniber}},
  \bibinfo {author} {\bibfnamefont {F.}~\bibnamefont {Flassig}}, \bibinfo
  {author} {\bibfnamefont {S.}~\bibnamefont {Lichtmannecker}}, \bibinfo
  {author} {\bibfnamefont {K.}~\bibnamefont {M{\"u}ller}}, \bibinfo {author}
  {\bibfnamefont {A.}~\bibnamefont {Andrejew}}, \bibinfo {author}
  {\bibfnamefont {J.}~\bibnamefont {Vuckovic}}, \bibinfo {author}
  {\bibfnamefont {R.}~\bibnamefont {Gross}}, \ and\ \bibinfo {author}
  {\bibfnamefont {J.}~\bibnamefont {Finley}},\ }\href {\doibase
  10.1021/acs.nanolett.5b01444} {\bibfield  {journal} {\bibinfo  {journal}
  {Nano Lett.}\ }\textbf {\bibinfo {volume} {15}},\ \bibinfo {pages} {5208}
  (\bibinfo {year} {2015})}\BibitemShut {NoStop}%
\bibitem [{\citenamefont {He}\ \emph {et~al.}(2019)\citenamefont {He},
  \citenamefont {Wang}, \citenamefont {Wang}, \citenamefont {Chen},
  \citenamefont {Ding}, \citenamefont {Qin}, \citenamefont {Duan},
  \citenamefont {Chen}, \citenamefont {Li}, \citenamefont {Liu}, \citenamefont
  {Schneider}, \citenamefont {Atat{\"u}re}, \citenamefont {H{\"o}fling},
  \citenamefont {Lu},\ and\ \citenamefont {Pan}}]{He:19}%
  \BibitemOpen
  \bibfield  {author} {\bibinfo {author} {\bibfnamefont {Y.-M.}\ \bibnamefont
  {He}}, \bibinfo {author} {\bibfnamefont {H.}~\bibnamefont {Wang}}, \bibinfo
  {author} {\bibfnamefont {C.}~\bibnamefont {Wang}}, \bibinfo {author}
  {\bibfnamefont {M.~C.}\ \bibnamefont {Chen}}, \bibinfo {author}
  {\bibfnamefont {X.}~\bibnamefont {Ding}}, \bibinfo {author} {\bibfnamefont
  {J.}~\bibnamefont {Qin}}, \bibinfo {author} {\bibfnamefont {Z.~C.}\
  \bibnamefont {Duan}}, \bibinfo {author} {\bibfnamefont {S.}~\bibnamefont
  {Chen}}, \bibinfo {author} {\bibfnamefont {J.~P.}\ \bibnamefont {Li}},
  \bibinfo {author} {\bibfnamefont {R.-Z.}\ \bibnamefont {Liu}}, \bibinfo
  {author} {\bibfnamefont {C.}~\bibnamefont {Schneider}}, \bibinfo {author}
  {\bibfnamefont {M.}~\bibnamefont {Atat{\"u}re}}, \bibinfo {author}
  {\bibfnamefont {S.}~\bibnamefont {H{\"o}fling}}, \bibinfo {author}
  {\bibfnamefont {C.-Y.}\ \bibnamefont {Lu}}, \ and\ \bibinfo {author}
  {\bibfnamefont {J.-W.}\ \bibnamefont {Pan}},\ }\href {\doibase
  10.1038/s41567-019-0585-6} {\bibfield  {journal} {\bibinfo  {journal} {Nat.
  Phys.}\ }\textbf {\bibinfo {volume} {15}},\ \bibinfo {pages} {941} (\bibinfo
  {year} {2019})}\BibitemShut {NoStop}%
\bibitem [{\citenamefont {Meitl}\ \emph {et~al.}(2006)\citenamefont {Meitl},
  \citenamefont {Zhu}, \citenamefont {Kumar}, \citenamefont {Lee},
  \citenamefont {Feng}, \citenamefont {Huang}, \citenamefont {Adesida},
  \citenamefont {Nuzzo},\ and\ \citenamefont {Rogers}}]{Meitl:06}%
  \BibitemOpen
  \bibfield  {author} {\bibinfo {author} {\bibfnamefont {M.~A.}\ \bibnamefont
  {Meitl}}, \bibinfo {author} {\bibfnamefont {Z.-T.}\ \bibnamefont {Zhu}},
  \bibinfo {author} {\bibfnamefont {V.}~\bibnamefont {Kumar}}, \bibinfo
  {author} {\bibfnamefont {K.~J.}\ \bibnamefont {Lee}}, \bibinfo {author}
  {\bibfnamefont {X.}~\bibnamefont {Feng}}, \bibinfo {author} {\bibfnamefont
  {Y.~Y.}\ \bibnamefont {Huang}}, \bibinfo {author} {\bibfnamefont
  {I.}~\bibnamefont {Adesida}}, \bibinfo {author} {\bibfnamefont {R.~G.}\
  \bibnamefont {Nuzzo}}, \ and\ \bibinfo {author} {\bibfnamefont {J.~A.}\
  \bibnamefont {Rogers}},\ }\href {\doibase 10.1038/nmat1532} {\bibfield
  {journal} {\bibinfo  {journal} {Nat. Mater.}\ }\textbf {\bibinfo {volume}
  {5}},\ \bibinfo {pages} {33} (\bibinfo {year} {2006})}\BibitemShut {NoStop}%
\bibitem [{\citenamefont {Kim}\ \emph {et~al.}(2011)\citenamefont {Kim},
  \citenamefont {Cho}, \citenamefont {Lee}, \citenamefont {Lee}, \citenamefont
  {Chae}, \citenamefont {Kim}, \citenamefont {Kim}, \citenamefont {Kwon},
  \citenamefont {Amaratunga}, \citenamefont {Lee}, \citenamefont {Choi},
  \citenamefont {Kuk}, \citenamefont {Kim},\ and\ \citenamefont
  {Kim}}]{Kim:11}%
  \BibitemOpen
  \bibfield  {author} {\bibinfo {author} {\bibfnamefont {T.-H.}\ \bibnamefont
  {Kim}}, \bibinfo {author} {\bibfnamefont {K.-S.}\ \bibnamefont {Cho}},
  \bibinfo {author} {\bibfnamefont {E.~K.}\ \bibnamefont {Lee}}, \bibinfo
  {author} {\bibfnamefont {S.~J.}\ \bibnamefont {Lee}}, \bibinfo {author}
  {\bibfnamefont {J.}~\bibnamefont {Chae}}, \bibinfo {author} {\bibfnamefont
  {J.~W.}\ \bibnamefont {Kim}}, \bibinfo {author} {\bibfnamefont {D.~H.}\
  \bibnamefont {Kim}}, \bibinfo {author} {\bibfnamefont {J.-Y.}\ \bibnamefont
  {Kwon}}, \bibinfo {author} {\bibfnamefont {G.}~\bibnamefont {Amaratunga}},
  \bibinfo {author} {\bibfnamefont {S.~Y.}\ \bibnamefont {Lee}}, \bibinfo
  {author} {\bibfnamefont {B.~L.}\ \bibnamefont {Choi}}, \bibinfo {author}
  {\bibfnamefont {Y.}~\bibnamefont {Kuk}}, \bibinfo {author} {\bibfnamefont
  {J.~M.}\ \bibnamefont {Kim}}, \ and\ \bibinfo {author} {\bibfnamefont
  {K.}~\bibnamefont {Kim}},\ }\href {\doibase 10.1038/nphoton.2011.12}
  {\bibfield  {journal} {\bibinfo  {journal} {Nat. Photon.}\ }\textbf {\bibinfo
  {volume} {5}},\ \bibinfo {pages} {176} (\bibinfo {year} {2011})}\BibitemShut
  {NoStop}%
\bibitem [{\citenamefont {Justice}\ \emph {et~al.}(2012)\citenamefont
  {Justice}, \citenamefont {Bower}, \citenamefont {Meitl}, \citenamefont
  {Mooney}, \citenamefont {Gubbins},\ and\ \citenamefont
  {Corbett}}]{Justice:12}%
  \BibitemOpen
  \bibfield  {author} {\bibinfo {author} {\bibfnamefont {J.}~\bibnamefont
  {Justice}}, \bibinfo {author} {\bibfnamefont {C.}~\bibnamefont {Bower}},
  \bibinfo {author} {\bibfnamefont {M.}~\bibnamefont {Meitl}}, \bibinfo
  {author} {\bibfnamefont {M.~B.}\ \bibnamefont {Mooney}}, \bibinfo {author}
  {\bibfnamefont {M.~A.}\ \bibnamefont {Gubbins}}, \ and\ \bibinfo {author}
  {\bibfnamefont {B.}~\bibnamefont {Corbett}},\ }\href {\doibase
  10.1038/nphoton.2012.204} {\bibfield  {journal} {\bibinfo  {journal} {Nat.
  Photon.}\ }\textbf {\bibinfo {volume} {6}},\ \bibinfo {pages} {610} (\bibinfo
  {year} {2012})}\BibitemShut {NoStop}%
\bibitem [{\citenamefont {Chakraborty}\ \emph {et~al.}(2020)\citenamefont
  {Chakraborty}, \citenamefont {Carolan}, \citenamefont {Clark}, \citenamefont
  {Bunandar}, \citenamefont {Gilbert}, \citenamefont {Notaros}, \citenamefont
  {Watts},\ and\ \citenamefont {Englund}}]{Chakraborty:20}%
  \BibitemOpen
  \bibfield  {author} {\bibinfo {author} {\bibfnamefont {U.}~\bibnamefont
  {Chakraborty}}, \bibinfo {author} {\bibfnamefont {J.}~\bibnamefont
  {Carolan}}, \bibinfo {author} {\bibfnamefont {G.}~\bibnamefont {Clark}},
  \bibinfo {author} {\bibfnamefont {D.}~\bibnamefont {Bunandar}}, \bibinfo
  {author} {\bibfnamefont {G.}~\bibnamefont {Gilbert}}, \bibinfo {author}
  {\bibfnamefont {J.}~\bibnamefont {Notaros}}, \bibinfo {author} {\bibfnamefont
  {M.~R.}\ \bibnamefont {Watts}}, \ and\ \bibinfo {author} {\bibfnamefont
  {D.~R.}\ \bibnamefont {Englund}},\ }\href {\doibase 10.1364/OPTICA.403178}
  {\bibfield  {journal} {\bibinfo  {journal} {Optica}\ }\textbf {\bibinfo
  {volume} {7}},\ \bibinfo {pages} {1385} (\bibinfo {year} {2020})}\BibitemShut
  {NoStop}%
\bibitem [{\citenamefont {Ferrari}\ \emph {et~al.}(2018)\citenamefont
  {Ferrari}, \citenamefont {Schuck},\ and\ \citenamefont
  {Pernice}}]{Ferrari:18}%
  \BibitemOpen
  \bibfield  {author} {\bibinfo {author} {\bibfnamefont {S.}~\bibnamefont
  {Ferrari}}, \bibinfo {author} {\bibfnamefont {C.}~\bibnamefont {Schuck}}, \
  and\ \bibinfo {author} {\bibfnamefont {W.}~\bibnamefont {Pernice}},\ }\href
  {\doibase 10.1515/nanoph-2018-0059} {\bibfield  {journal} {\bibinfo
  {journal} {Nanophotonics}\ }\textbf {\bibinfo {volume} {7}},\ \bibinfo
  {pages} {1725} (\bibinfo {year} {2018})}\BibitemShut {NoStop}%
\bibitem [{\citenamefont {Gyger}\ \emph {et~al.}(2021)\citenamefont {Gyger},
  \citenamefont {Zichi}, \citenamefont {Schweickert}, \citenamefont {Elshaari},
  \citenamefont {Steinhauer}, \citenamefont {Covre~da Silva}, \citenamefont
  {Rastelli}, \citenamefont {Zwiller}, \citenamefont {J{\"o}ns},\ and\
  \citenamefont {Errando-Herranz}}]{Gyger:21}%
  \BibitemOpen
  \bibfield  {author} {\bibinfo {author} {\bibfnamefont {S.}~\bibnamefont
  {Gyger}}, \bibinfo {author} {\bibfnamefont {J.}~\bibnamefont {Zichi}},
  \bibinfo {author} {\bibfnamefont {L.}~\bibnamefont {Schweickert}}, \bibinfo
  {author} {\bibfnamefont {A.~W.}\ \bibnamefont {Elshaari}}, \bibinfo {author}
  {\bibfnamefont {S.}~\bibnamefont {Steinhauer}}, \bibinfo {author}
  {\bibfnamefont {S.~F.}\ \bibnamefont {Covre~da Silva}}, \bibinfo {author}
  {\bibfnamefont {A.}~\bibnamefont {Rastelli}}, \bibinfo {author}
  {\bibfnamefont {V.}~\bibnamefont {Zwiller}}, \bibinfo {author} {\bibfnamefont
  {K.~D.}\ \bibnamefont {J{\"o}ns}}, \ and\ \bibinfo {author} {\bibfnamefont
  {C.}~\bibnamefont {Errando-Herranz}},\ }\href {\doibase
  10.1038/s41467-021-21624-3} {\bibfield  {journal} {\bibinfo  {journal} {Nat.
  Commun.}\ }\textbf {\bibinfo {volume} {12}},\ \bibinfo {pages} {1408}
  (\bibinfo {year} {2021})}\BibitemShut {NoStop}%
\bibitem [{\citenamefont {Rudolph}(2017)}]{Rudolph:17}%
  \BibitemOpen
  \bibfield  {author} {\bibinfo {author} {\bibfnamefont {T.}~\bibnamefont
  {Rudolph}},\ }\href {\doibase 10.1063/1.4976737} {\bibfield  {journal}
  {\bibinfo  {journal} {APL Photonics}\ }\textbf {\bibinfo {volume} {2}},\
  \bibinfo {pages} {030901} (\bibinfo {year} {2017})}\BibitemShut {NoStop}%
\bibitem [{\citenamefont {Steinbrecher}\ \emph {et~al.}(2019)\citenamefont
  {Steinbrecher}, \citenamefont {Olson}, \citenamefont {Englund},\ and\
  \citenamefont {Carolan}}]{Steinbrecher:19}%
  \BibitemOpen
  \bibfield  {author} {\bibinfo {author} {\bibfnamefont {G.~R.}\ \bibnamefont
  {Steinbrecher}}, \bibinfo {author} {\bibfnamefont {J.~P.}\ \bibnamefont
  {Olson}}, \bibinfo {author} {\bibfnamefont {D.}~\bibnamefont {Englund}}, \
  and\ \bibinfo {author} {\bibfnamefont {J.}~\bibnamefont {Carolan}},\ }\href
  {\doibase 10.1038/s41534-019-0174-7} {\bibfield  {journal} {\bibinfo
  {journal} {npj Quantum Inf.}\ }\textbf {\bibinfo {volume} {5}},\ \bibinfo
  {pages} {60} (\bibinfo {year} {2019})}\BibitemShut {NoStop}%
\bibitem [{\citenamefont {Lopez-Pastor}\ and\ \citenamefont
  {Marquardt}(2021)}]{Lopez:21}%
  \BibitemOpen
  \bibfield  {author} {\bibinfo {author} {\bibfnamefont {V.}~\bibnamefont
  {Lopez-Pastor}}\ and\ \bibinfo {author} {\bibfnamefont {F.}~\bibnamefont
  {Marquardt}},\ }\href {\doibase 10.48550/arXiv.2103.04992} {\ ,\ \bibinfo
  {pages} {Preprint at https://arxiv.org/abs/2103.04992} (\bibinfo {year}
  {2021})}\BibitemShut {NoStop}%
\bibitem [{\citenamefont {Liu}\ \emph {et~al.}(2018)\citenamefont {Liu},
  \citenamefont {Konthasinghe}, \citenamefont {Davan\ifmmode~\mbox{\c{c}}\else
  \c{c}\fi{}o}, \citenamefont {Lawall}, \citenamefont {Anant}, \citenamefont
  {Verma}, \citenamefont {Mirin}, \citenamefont {Nam}, \citenamefont {Song},
  \citenamefont {Ma}, \citenamefont {Chen}, \citenamefont {Ni}, \citenamefont
  {Niu},\ and\ \citenamefont {Srinivasan}}]{Liu:18}%
  \BibitemOpen
  \bibfield  {author} {\bibinfo {author} {\bibfnamefont {J.}~\bibnamefont
  {Liu}}, \bibinfo {author} {\bibfnamefont {K.}~\bibnamefont {Konthasinghe}},
  \bibinfo {author} {\bibfnamefont {M.}~\bibnamefont
  {Davan\ifmmode~\mbox{\c{c}}\else \c{c}\fi{}o}}, \bibinfo {author}
  {\bibfnamefont {J.}~\bibnamefont {Lawall}}, \bibinfo {author} {\bibfnamefont
  {V.}~\bibnamefont {Anant}}, \bibinfo {author} {\bibfnamefont
  {V.}~\bibnamefont {Verma}}, \bibinfo {author} {\bibfnamefont
  {R.}~\bibnamefont {Mirin}}, \bibinfo {author} {\bibfnamefont {S.~W.}\
  \bibnamefont {Nam}}, \bibinfo {author} {\bibfnamefont {J.~D.}\ \bibnamefont
  {Song}}, \bibinfo {author} {\bibfnamefont {B.}~\bibnamefont {Ma}}, \bibinfo
  {author} {\bibfnamefont {Z.~S.}\ \bibnamefont {Chen}}, \bibinfo {author}
  {\bibfnamefont {H.~Q.}\ \bibnamefont {Ni}}, \bibinfo {author} {\bibfnamefont
  {Z.~C.}\ \bibnamefont {Niu}}, \ and\ \bibinfo {author} {\bibfnamefont
  {K.}~\bibnamefont {Srinivasan}},\ }\href {\doibase
  10.1103/PhysRevApplied.9.064019} {\bibfield  {journal} {\bibinfo  {journal}
  {Phys. Rev. Appl.}\ }\textbf {\bibinfo {volume} {9}},\ \bibinfo {pages}
  {064019} (\bibinfo {year} {2018})}\BibitemShut {NoStop}%
\bibitem [{\citenamefont {Schnauber}\ \emph {et~al.}(2019)\citenamefont
  {Schnauber}, \citenamefont {Singh}, \citenamefont {Schall}, \citenamefont
  {Park}, \citenamefont {Song}, \citenamefont {Rodt}, \citenamefont
  {Srinivasan}, \citenamefont {Reitzenstein},\ and\ \citenamefont
  {Davanco}}]{Schnauber:19}%
  \BibitemOpen
  \bibfield  {author} {\bibinfo {author} {\bibfnamefont {P.}~\bibnamefont
  {Schnauber}}, \bibinfo {author} {\bibfnamefont {A.}~\bibnamefont {Singh}},
  \bibinfo {author} {\bibfnamefont {J.}~\bibnamefont {Schall}}, \bibinfo
  {author} {\bibfnamefont {S.~I.}\ \bibnamefont {Park}}, \bibinfo {author}
  {\bibfnamefont {J.~D.}\ \bibnamefont {Song}}, \bibinfo {author}
  {\bibfnamefont {S.}~\bibnamefont {Rodt}}, \bibinfo {author} {\bibfnamefont
  {K.}~\bibnamefont {Srinivasan}}, \bibinfo {author} {\bibfnamefont
  {S.}~\bibnamefont {Reitzenstein}}, \ and\ \bibinfo {author} {\bibfnamefont
  {M.}~\bibnamefont {Davanco}},\ }\href {\doibase 10.1021/acs.nanolett.9b02758}
  {\bibfield  {journal} {\bibinfo  {journal} {Nano Lett.}\ }\textbf {\bibinfo
  {volume} {19}},\ \bibinfo {pages} {7164} (\bibinfo {year}
  {2019})}\BibitemShut {NoStop}%
\bibitem [{\citenamefont {Panuski}\ \emph {et~al.}(2022)\citenamefont
  {Panuski}, \citenamefont {Christen}, \citenamefont {Minkov}, \citenamefont
  {Brabec}, \citenamefont {Trajtenberg-Mills}, \citenamefont {Griffiths},
  \citenamefont {McKendry}, \citenamefont {Leake}, \citenamefont {Coleman},
  \citenamefont {Tran}, \citenamefont {St~Louis}, \citenamefont {Mucci},
  \citenamefont {Horvath}, \citenamefont {Westwood-Bachman}, \citenamefont
  {Preble}, \citenamefont {Dawson}, \citenamefont {Strain}, \citenamefont
  {Fanto},\ and\ \citenamefont {Englund}}]{Panuski:22}%
  \BibitemOpen
  \bibfield  {author} {\bibinfo {author} {\bibfnamefont {C.~L.}\ \bibnamefont
  {Panuski}}, \bibinfo {author} {\bibfnamefont {I.}~\bibnamefont {Christen}},
  \bibinfo {author} {\bibfnamefont {M.}~\bibnamefont {Minkov}}, \bibinfo
  {author} {\bibfnamefont {C.~J.}\ \bibnamefont {Brabec}}, \bibinfo {author}
  {\bibfnamefont {S.}~\bibnamefont {Trajtenberg-Mills}}, \bibinfo {author}
  {\bibfnamefont {A.~D.}\ \bibnamefont {Griffiths}}, \bibinfo {author}
  {\bibfnamefont {J.~J.~D.}\ \bibnamefont {McKendry}}, \bibinfo {author}
  {\bibfnamefont {G.~L.}\ \bibnamefont {Leake}}, \bibinfo {author}
  {\bibfnamefont {D.~J.}\ \bibnamefont {Coleman}}, \bibinfo {author}
  {\bibfnamefont {C.}~\bibnamefont {Tran}}, \bibinfo {author} {\bibfnamefont
  {J.}~\bibnamefont {St~Louis}}, \bibinfo {author} {\bibfnamefont
  {J.}~\bibnamefont {Mucci}}, \bibinfo {author} {\bibfnamefont
  {C.}~\bibnamefont {Horvath}}, \bibinfo {author} {\bibfnamefont {J.~N.}\
  \bibnamefont {Westwood-Bachman}}, \bibinfo {author} {\bibfnamefont {S.~F.}\
  \bibnamefont {Preble}}, \bibinfo {author} {\bibfnamefont {M.~D.}\
  \bibnamefont {Dawson}}, \bibinfo {author} {\bibfnamefont {M.~J.}\
  \bibnamefont {Strain}}, \bibinfo {author} {\bibfnamefont {M.~L.}\
  \bibnamefont {Fanto}}, \ and\ \bibinfo {author} {\bibfnamefont {D.~R.}\
  \bibnamefont {Englund}},\ }\href {\doibase 10.1038/s41566-022-01086-9}
  {\bibfield  {journal} {\bibinfo  {journal} {Nat. Photon.}\ }\textbf {\bibinfo
  {volume} {16}},\ \bibinfo {pages} {834} (\bibinfo {year} {2022})}\BibitemShut
  {NoStop}%
\end{thebibliography}
\end{document}


\title{Supplementary Document for: \\
Tunable quantum  emitters integrated on large-scale foundry silicon photonics}

\author{Hugo~Larocque}
\affiliation{Research Laboratory of Electronics, Massachusetts Institute of Technology, Cambridge, Massachusetts 02139, USA}
\email{hlarocqu@mit.edu}

\author{Mustafa~Atabey~Buyukkaya}
\affiliation{Department of Electrical and Computer Engineering and Institute for Research in Electronics and Applied Physics, University of Maryland, College Park, Maryland 20742, USA}

\author{Carlos~Errando-Herranz}
\affiliation{Research Laboratory of Electronics, Massachusetts Institute of Technology, Cambridge, Massachusetts 02139, USA}
\affiliation{Institute of Physics, University of M\"unster, 48149, M\"unster, Germany}

\author{Samuel~Harper}
\affiliation{Department of Electrical and Computer Engineering and Institute for Research in Electronics and Applied Physics, University of Maryland, College Park, Maryland 20742, USA}

\author{Jacques~Carolan}
\affiliation{Research Laboratory of Electronics, Massachusetts Institute of Technology, Cambridge, Massachusetts 02139, USA}
\affiliation{Present Address: Wolfson Institute for Biomedical Research, University College London, London, UK}

\author{Chang-Min~Lee}
\affiliation{Department of Electrical and Computer Engineering and Institute for Research in Electronics and Applied Physics, University of Maryland, College Park, Maryland 20742, USA}

\author{Christopher~J.~K.~Richardson}
\affiliation{Laboratory for Physical Sciences, University of Maryland, College Park, Maryland 20740, USA}

\author{Gerald~L.~Leake}
\affiliation{State University of New York Polytechnic Institute, Albany, New York 12203, USA}

\author{Daniel~J.~Coleman}
\affiliation{State University of New York Polytechnic Institute, Albany, New York 12203, USA}

\author{Michael~L.~Fanto}
\affiliation{Air Force Research Laboratory, Information Directorate, Rome, New York, 13441, USA}

\author{Edo~Waks}
\affiliation{Department of Electrical and Computer Engineering and Institute for Research in Electronics and Applied Physics, University of Maryland, College Park, Maryland 20742, USA}

\author{Dirk~Englund}
\affiliation{Research Laboratory of Electronics, Massachusetts Institute of Technology, Cambridge, Massachusetts 02139, USA}

\begin{abstract}
\end{abstract}

\maketitle

\section{Photoluminescence spectrum repeatability}

\begin{figure}[h!]
    \centering
    \includegraphics[width=0.8\linewidth]{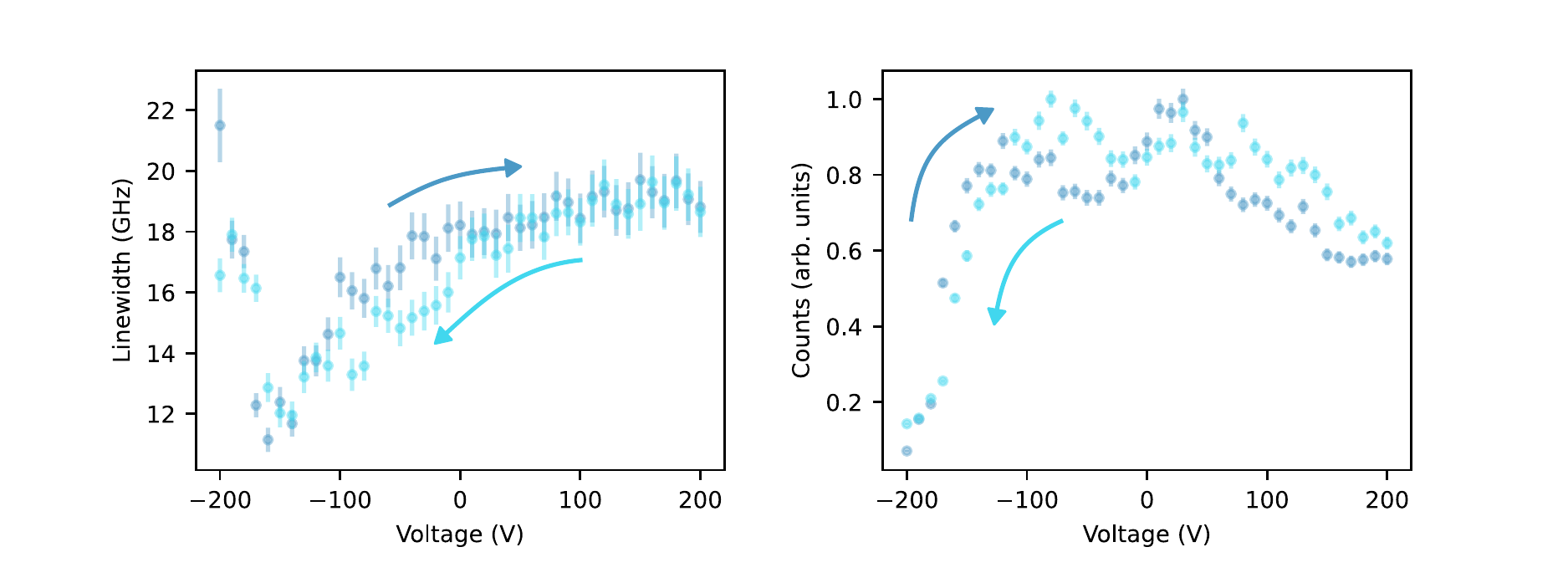}
    \caption{\textbf{Linewidth and brightness of E2 with applied voltage.} The values are extracted from the optimal parameters for fits of measured spectra to Lorentzian lineshapes. Error bars correspond to one standard deviation of the fitted parameter (see Methods). We provide data taken over two voltage scans. The direction of the arrows included in the plots provide the direction of the voltage scan for its correspondingly colored data set.}
    \label{fig:E2emission}
\end{figure}

Supplementary Figure~\ref{fig:E2emission}, plots the extracted linewidth and brightness of emitter E2 discussed in the main text. We extract these quantities by fitting the emission spectra of the emitter under above-band excitation with a CW laser with a wavelength of 780~nm acquired while we sweep the voltage from -200~V to 200~V across our tuning device. Figure~4a from the main text displays some of these spectra. We observe that the extracted quantities from two sweeps are very similar for a given voltage. We attribute minor discrepancies to factors such as drift in the position of our excitation beam and the finite resolution of our specrometer, which is around 7~GHz.

\section{Emitter properties under pulsed excitation}

\begin{figure}[h!]
    \centering
    \includegraphics[width=\linewidth]{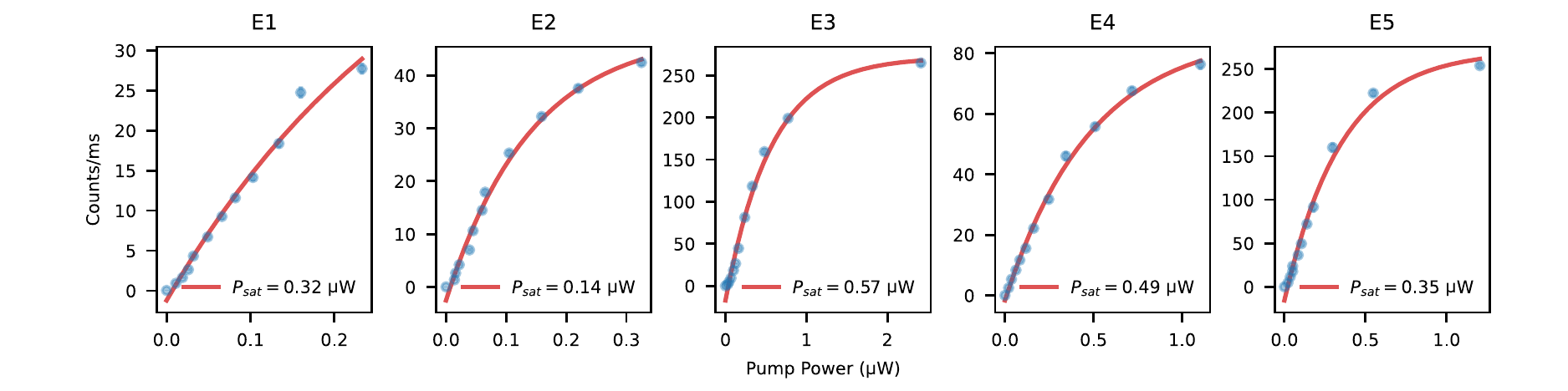}
    \caption{\textbf{Saturation measurements of emitters along the InP waveguide.} Saturation curves of various emitters along a transferred chiplet waveguide under pulsed above-band excitation at a wavelength of 776~nm using a NA=0.55 objective. We also provide fits of the data to $I(P)=I_\text{sat}(1-\exp(-P/P_\text{E1,sat}))$ and include the fitted $P_\text{sat}$ in each emitter's subpanel.}
    \label{fig:suppSat}
\end{figure}

\begin{figure}[h!]
    \centering
    \includegraphics[width=\linewidth]{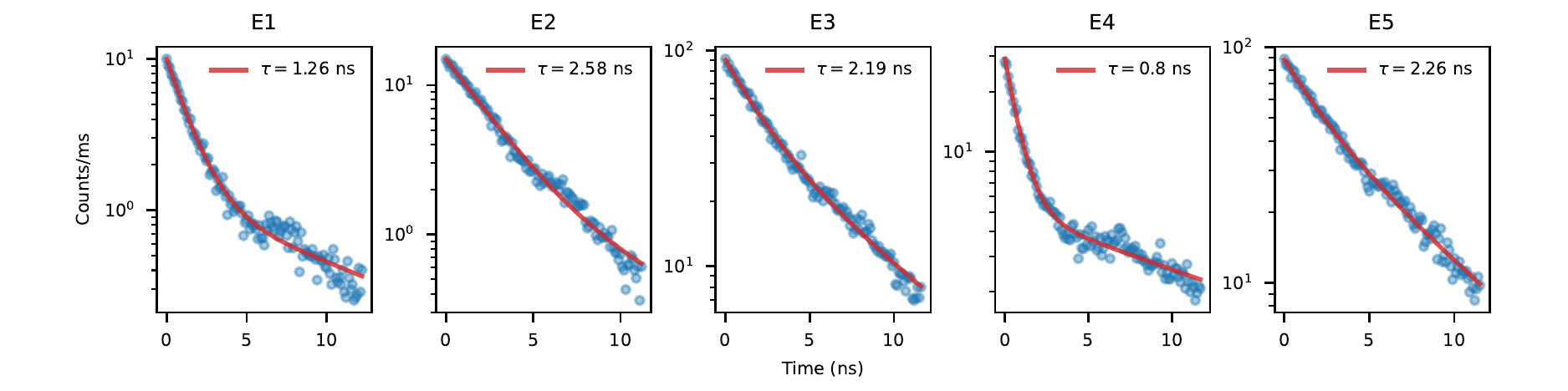}
    \caption{\textbf{Lifetime measurements of emitters along the InP waveguide.} Radiative decay time trace of various emitters along a transferred chiplet waveguide. We also provide fits of the data to a double-exponential decay function $I(t)=a\exp(-t/\tau)+\tilde{a}\exp(-t/\tilde{\tau})$, where $a>\tilde{a}$. We include the best fit value for $\tau$ of each emitter in their respective subpanels. We collected the data with the same equipment used in Supplementary Figure~\ref{fig:suppSat} and a third of the maximum pump power considered in the corresponding saturation measurement.}
    \label{fig:suppLife}
\end{figure}

Here, we report the saturation powers and lifetimes of other emitters under above-band excitation using a 776~nm pulsed laser with a pulse width of 120~fs and a 80~MHz repetition rate. Supplementary Figure~\ref{fig:suppSat} provides the saturation curve for five different emitters, E1-E5, whereas Supplementary Figure~\ref{fig:suppLife} plots their corresponding radiative decay time traces. We observe that the saturation power, $P_\text{sat}$, of our emitters ranges from 0.14~\textmu W to 0.57~\textmu W, whereas the dominant lifetime, $\tau$, varies within 0.8~ns to 2.58~ns. The influence of a second time constant, $\tilde\tau$, is more noticeable in emitters with shorter lifetimes, e.g. E1 and E4, thereby suggesting the presence of an adverse carrier recombination process to excitonic recombination in the quantum dots.

\section{Transfer Accuracy}

We estimate the accuracy of our chiplet placement method by examining 44 transferred chiplets distributed over 4 chips. We visually locate the extremities of the integrated InP and native Si waveguides of each hybrid device on optical micrographs collected with a NA=0.95 microscope objective. We then convert the resulting pixel values attributed to these extremities to absolute distance values. We perform this conversion with a factor extracted from 65 images of a known feature size on our device. Specifically, this distance corresponds to the width of the chiplet transfer pad, which is known to be 10~\textmu m. We found the pixel distance attributed to this feature to be $274 \pm 8$ pixels, hence a conversion factor of 36~nm/pixel. Based on the standard deviation of this value, the error attributed to manually delimiting this length was 294~nm, which roughly corresponds to the diffraction limit of our microscope system. Therefore, we attribute an uncertainty of 294~nm to the extracted waveguide extremity positions.

Based on these positions, we extract the  relative rotation and displacement between the InP and Si waveguides. Here, we define displacement as the distance from the Si waveguide to the tip of the InP taper. We provide these values in Supplementary Figure~\ref{fig:histo}. After carrying the error through our 44 images, we obtain an average rotation and displacement of $0.03 \pm 0.13^\text{o}$ and $-15 \pm 44$~nm, respectively.

\begin{figure}[h!]
    \centering
    \includegraphics[width=\linewidth]{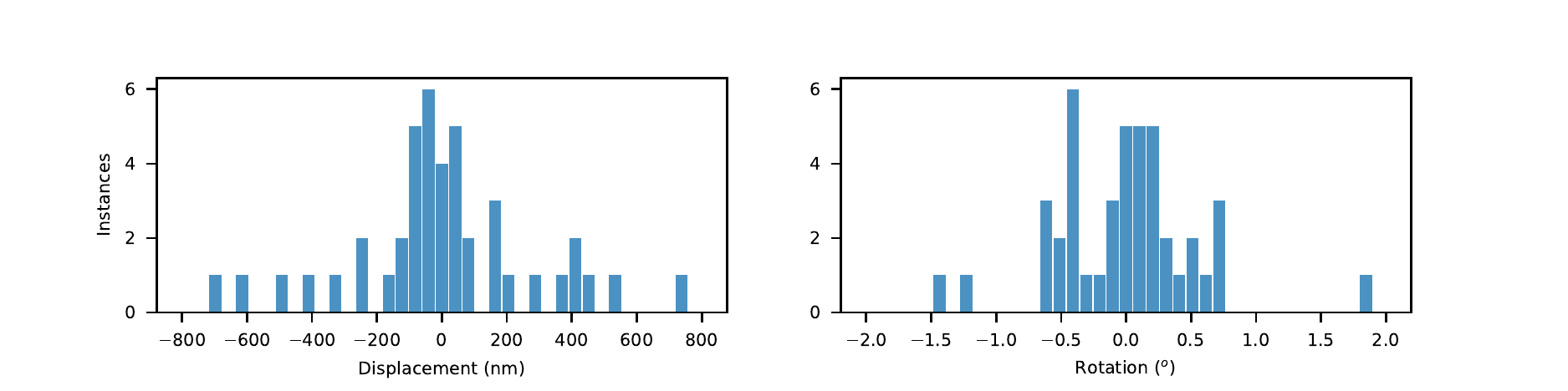}
    \caption{\textbf{Placement accuracy.} Displacement and rotation between the InP and Si waveguides of 44 fabricated hybrid devices.}
    \label{fig:histo}
\end{figure}

\section{Experimental chiplet to PIC coupling}

\begin{figure}[h!]
    \centering
    \includegraphics[width=\linewidth]{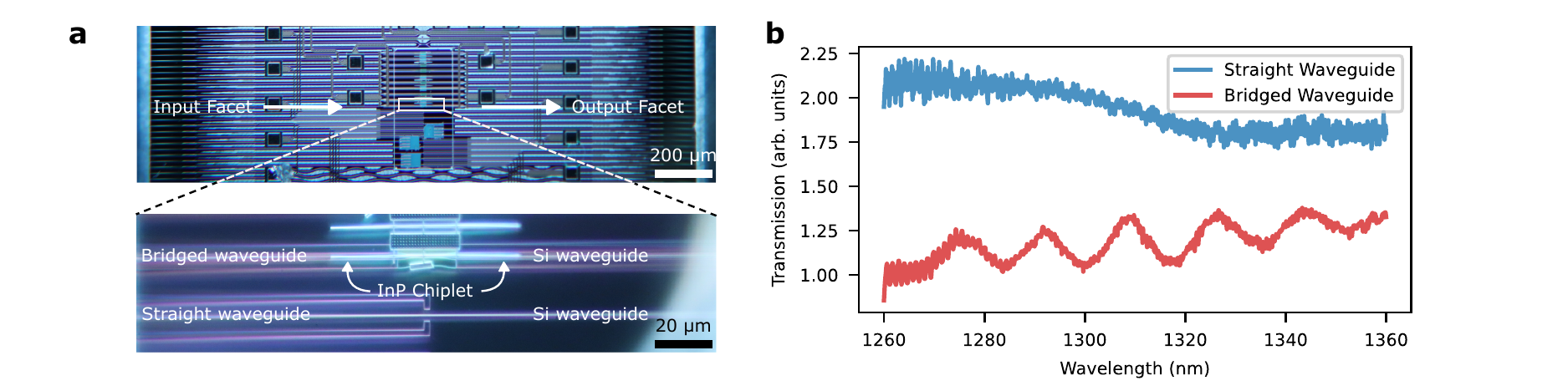}
    \caption{\textbf{Chiplet to PIC coupling.} \textbf{a,} Dark field optical micrographs of the structure for gauging chiplet-to-PIC coupling effiency. An InP chiplet bridges a gap between two Si adiabatic tapers to compare its transmission with a nearby straight silicon waveguide. Both the bridged and straight waveguide use the same edge couplers for fiber-to-waveguide facet coupling. \textbf{b,} Transmission through the straight and bridged waveguides.}
    \label{fig:coupling}
\end{figure}

We obtain the $86\%$ chiplet-to-PIC coupling metric reported in the main text by bridging a gap between two tapered Si waveguides with an InP chiplet. As shown in Supplementary Figure~\ref{fig:coupling}a, the chiplet consists of a waveguide where both ends are tapered, thereby enabling incoming light to cross the Si waveguide gap by means of optical coupling through the InP chiplet. Supplementary Figure~\ref{fig:coupling}b plots the transmission spectrum of the bridged structure and of an adjacent straight waveguide. Both structures are 2~mm long and rely on edge couplers for fiber-to-waveguide facet transmission. Optical coupling through the chiplet is optimal within wavelengths of 1340~nm and 1360~nm where we observe throughputs of 75\% of the straight waveguide transmission, thereby suggesting a $(0.75)^{1/2}= 86\%$ coupling efficiency between the chiplet and the PIC.

\section{Resonant pump rejection of the PIC}

We provide the raw photoluminescence excitation (PLE) data shown in Fig.~3aiii of the main text in Supplementary Figure~\ref{fig:ple}. Under the  10~\textmu W pump power conditions reported in the main text, our waveguide features good pump rejection as indicated by the resonance fluorescence being roughly 45 times greater than the scattered laser signal. A tenfold increase in pump power drastically reduces this factor as the emitter reaches saturation and more laser power scatters into the PIC.  

\begin{figure}[h!]
    \centering
    \includegraphics[width=0.5\linewidth]{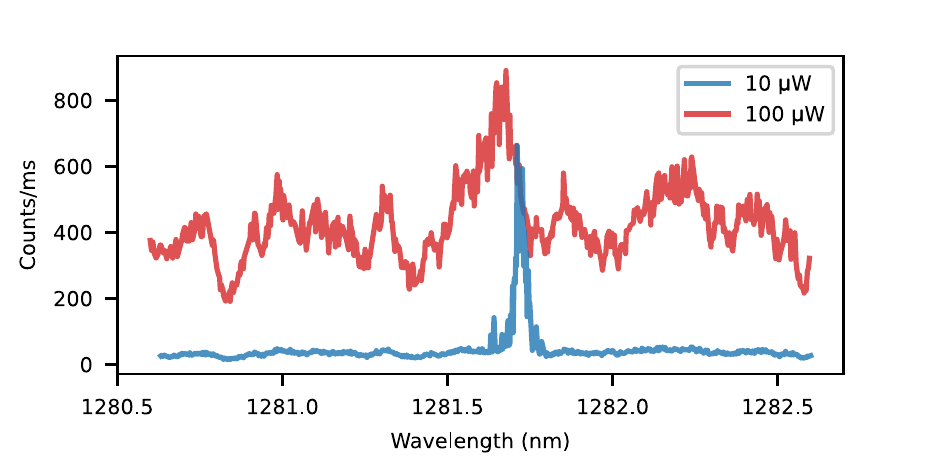}
    \caption{Raw photoluminescence spectra while resonantly pumping the quantum dot with 10~\textmu W and 100~\textmu W of optical power.}
    \label{fig:ple}
\end{figure}

\section{Pulsed Laser Reflections}

We attribute the spurious peaks found in the pulsed autocorrelation measurement of Fig.~3b(i) of the main text directly to the output of our pulsed laser. We provide a correlation measurement between the optical output of the laser and an electrical trigger from a fast photodiode in Supplementary Figure~\ref{fig:mira}. In addition to the main pulses separated by roughly 13~ns, we observe secondary pulses that are most likely due to reflections.

\begin{figure}[h!]
    \centering
    \includegraphics[width=\linewidth]{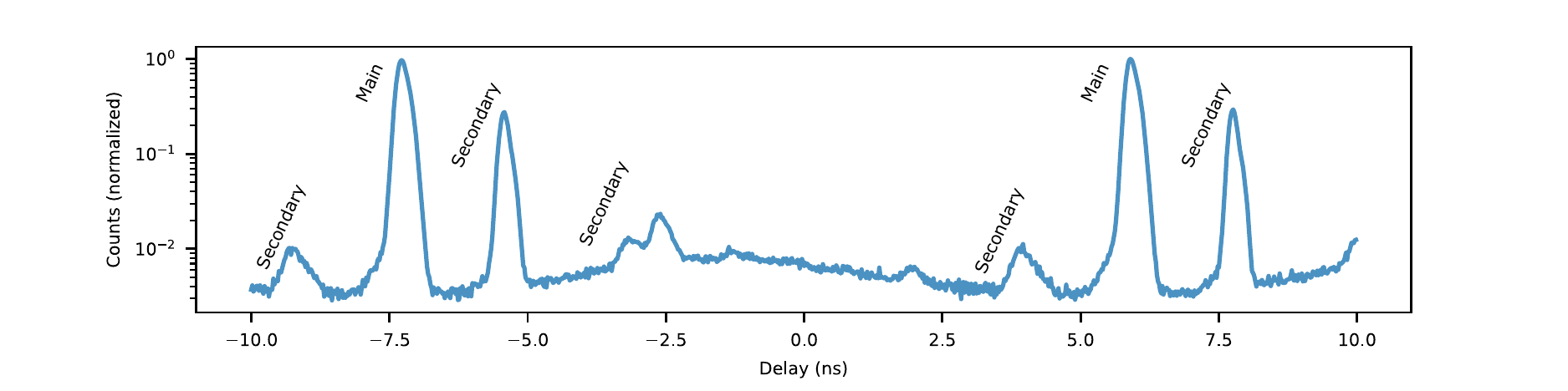}
    \caption{\textbf{Pulsed laser output.} Autocorrelation measurement between the optical and electrical trigger outputs of the laser used in our pulsed excitation experiments.}
    \label{fig:mira}
\end{figure}

\section{Local Material Variations}

\begin{figure}[h!]
    \centering
    \includegraphics[width=\linewidth]{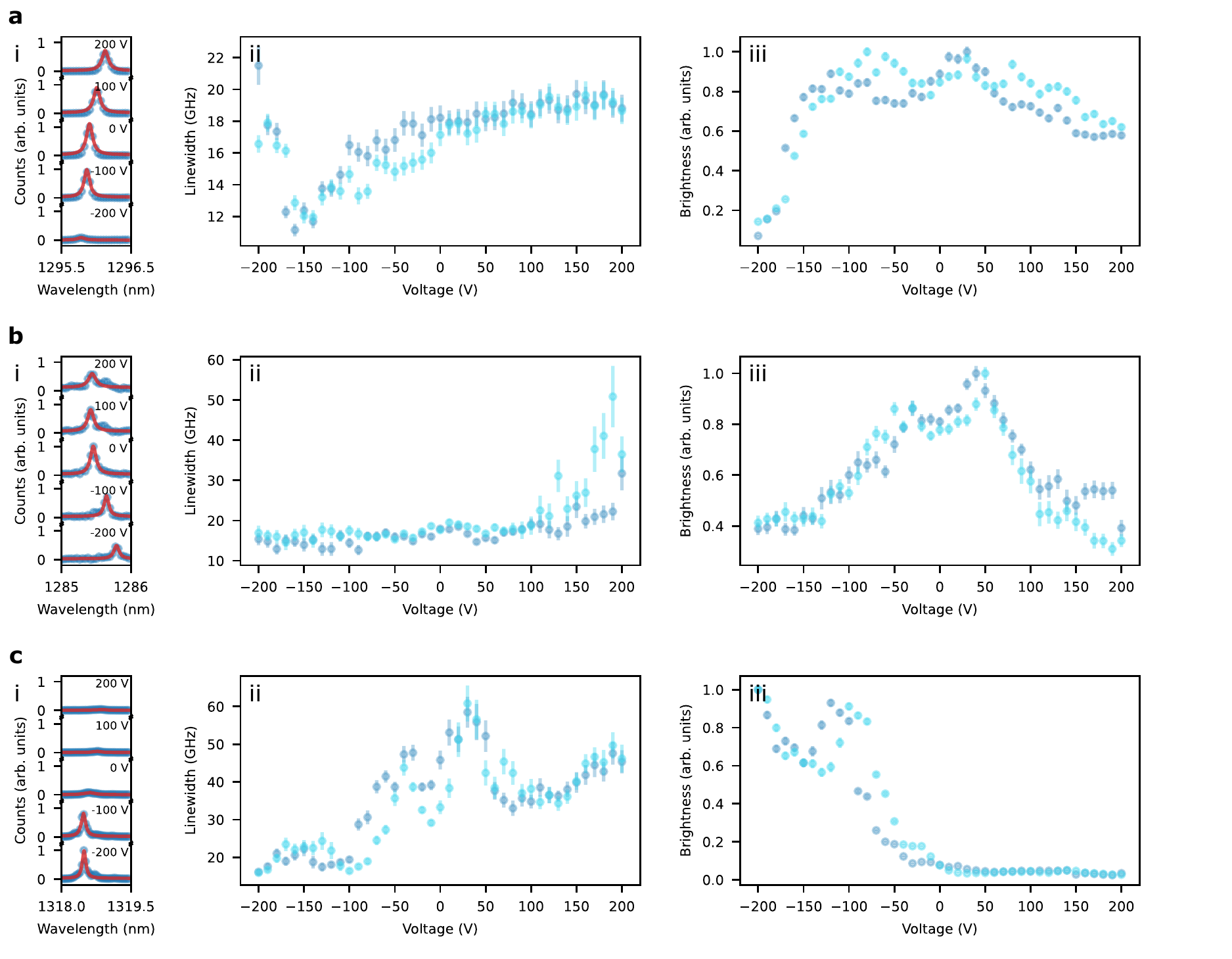}
    \caption{\textbf{Local material variations.} Applied voltage dependence of the emission properties of three emitters featuring dimming at \textbf{a,} low, \textbf{b,} low and high, and \textbf{c,} high applied voltages. The properties illustrated here include i, the emission line of the emitter at five different voltages, ii, the applied voltage dependence of the emitter linewidth, and iii, the applied voltage dependence of the emitter's relative brightness. The linewidth and relative brightness values correspond to the optimal parameters for fits of measured spectra to Lorentzian lineshapes. Error bars correspond to one standard deviation of the fitted parameter (see Methods). Data sets attributed to different voltage sweeps are correspondingly colored. }
    \label{fig:material}
\end{figure}

The trends in linewidth broadening and dimming exhibited by E2 shown in Fig.~4a of the main text are not universal to all emitters. Though E2 experiences these phenomena at low voltages, i.e. a large reverse bias as displayed in Supplementary Figure~\ref{fig:material}a, others can experience them at large biases in general or only at a large forward bias, as shown in Supplementary Figures~\ref{fig:material}b,c, respectively.

Prior work reports similar observations. Though local strain can have significant influence on the emission properties of quantum dots~\cite{Grim:19}, it can also influence their response to Stark tuning~\cite{Fry:00, Ostapenko:10}. Inhomogeneous stress imparted by our PECVD oxide buffer onto our transferred chiplets could justify similar behavior observed in our own samples.

\bibliographystyle{naturemag}
\bigbreak
\def\bibsection{}  
\noindent\textbf{Supplementary References}
\bigbreak